\documentclass[twocolumn]{aastex63}
\usepackage{natbib, color, amsmath}
\graphicspath{{./figures}}
\renewcommand{\vec}[1]{\boldsymbol{#1}}

\begin{document}

\title{Observational Constraints on the Radial Evolution of O$^{6+}$ Temperature and Differential Flow in the Inner Heliosphere}
\newcommand{\CfA}{\affiliation{Center for Astrophysics $\vert$ Harvard \& Smithsonian, 60 Garden Street, Cambridge, MA 02138, USA}}

\newcommand{\UA}{\affiliation{Lunar and Planetary Laboratory, University of Arizona, Tucson, AZ 85719, USA}}

\newcommand{\umich}{\affiliation{Department of Climate and Space Sciences and Engineering, The University of Michigan, Ann Arbor, MI, USA}}

\newcommand{\GSFC}{\affiliation{Heliophysics Science Division, NASA Goddard Space Flight Center, Greenbelt, MD 20771, USA}}

\newcommand{\MSSL}{\affiliation{Mullard Space Science Laboratory, University College London, Holmbury St. Mary, Dorking, Surrey, RH5 6NT, UK}}

\newcommand{\ICL}{\affiliation{Department of Physics, Imperial College London, London, SW7 2BW, UK}}

\newcommand{\GSFCSSED}{\affiliation{Solar System Exploration Division, NASA Goddard Space Flight Center, Greenbelt, MD 20771, USA}}

\newcommand{\UMBC}{\affiliation{University of Maryland, Baltimore County, Baltimore, MD 21250, USA}}

\newcommand{\SSL}{\affiliation{Space Sciences Laboratory, University of California, Berkeley, CA 94720, USA}}

\author[0000-0002-8748-2123]{Yeimy J. Rivera}
\CfA

\author[0000-0001-6038-1923]{Kristopher G. Klein}
\UA

\author[0009-0008-4095-9175]{Joseph H. Wang}
\ICL

\author[0000-0002-6276-7771]{Lorenzo Matteini}
\ICL

\author[0000-0002-0497-1096]{Daniel Verscharen}
\MSSL

\author[0000-0002-2576-0992]{Jesse T. Coburn}
\ICL

\author[0000-0002-6145-436X]{Samuel T. Badman}
\CfA

\author[0000-0003-1611-227X]{Susan T. Lepri}
\umich

\author[0000-0003-4437-0698]{Ryan M. Dewey}
\umich

\author[0000-0001-5956-9523]{Jim M. Raines}
\umich

\author[0000-0001-6673-3432]{B. L. Alterman}
\GSFC

\author[0000-0002-5524-645X]{Timothy J. Stubbs}
\GSFCSSED

\author[0000-0002-9280-0480]{Kevin C. Delano}
\GSFC\UMBC

\author[0000-0002-0396-0547]{Roberto Livi}
\SSL

\author[0000-0002-4149-7311]{Stefano A. Livi}
\affiliation{Southwest Research Institute, San Antonio, TX 78228, USA}
\umich

\author[0000-0003-3752-5700]{Antoinette B. Galvin}
\affiliation{University of New Hampshire, Durham, NH, USA}

\author[0000-0002-5982-4667]{Christopher J. Owen}
\MSSL

\author[0000-0002-7728-0085]{Michael L. Stevens}
\CfA

\begin{abstract}
Over decades of solar wind observations, heavy ions have been observed to have a higher temperature and flow faster than protons in the solar corona and heliosphere. Remote observations have largely been limited to the low corona ($< 4R_{\sun}$), while in situ observations for heavy ions ($Z>2$) have only been sampled at 1 au and beyond. As a result, theories that address heavy ion heating and acceleration remain largely unconstrained. With the launch of Solar Orbiter, heavy ion kinetics can be probed closer to the Sun, as close as the orbit of Mercury ($65R_{\sun}$), to examine their radial behavior. Through a statistical analysis of O$^{6+}$, this work provides a comprehensive analysis of the velocity and temperature of O$^{6+}$ from 0.3 au to 1 au. The study finds that the O$^{6+}$ relative drift, normalized to the local Alfv\'en speed, and its temperature compared to protons, both decrease with distance from the Sun and show some speed dependence. The O$^{6+}$ temperature is well fit by a single temperature adiabatic profile across all wind speeds, suggesting there is no significant heating at these heliocentric distances. This is in contrast to what is observed for protons and He$^{2+}$. Alfv\'enic fluctuations, with full 180$^{\circ}$ field rotation, create momentary negative differential streaming where the speed of O$^{6+}$ trails the protons.  The amount of negative differential streaming gradually increases at larger distances. These results provide critical constraints to the proposed mechanisms seeking to describe ion heating and acceleration in the solar wind.

\end{abstract}


\section{Introduction} \label{sec:intro}

The Sun creates a continuous stream of out-flowing plasma called the solar wind. The solar wind quickly expands as it escapes the magnetically dominated corona to fill the heliosphere. As the plasma expands, Coulomb collisions become less frequent as it transitions into a hot, collisionless plasma. Low in the corona, Coulomb collisions regulate particle distributions by continuously redistributing energy among the ions and electrons which maintain isotropic, Maxwellian velocity distributions. When collisions become less frequent, as the solar wind becomes more and more tenuous, non-thermal particle distributions are observed. These non-thermal distributions can be identified by high-energy ion and electron tails and beams, temperature anisotropies with respect to the background magnetic field, and differential heavy ion flows \citep{Marsch1982b, Marsch2006, Alterman2018, Verscharen2019_kinetics_review, Verniero2020}. The competition of collisions, waves/instabilities, and ongoing turbulence processes are responsible for the energy exchange and partitioning among ions and electrons, which can drive the heating and acceleration observed in the corona and throughout the inner heliosphere \citep{Tu_Marsch_1995, Gonzalez2021, Adhikari2021, Sioulas2022, Bandyopadhyay2023, Halekas2023, Shankarappa:2024, Rivera2024, Bourouaine2024, Rivera2025, Silwal2025}. 

In collisional thermodynamic equilibrium, the ions and electrons are described by a single temperature and flow speed. However, deviations from equilibrium are often observed, even in the corona, as non-thermal features develop as close as 1 R$_{\sun}$ (solar radii) from the solar surface in coronal holes. The UltraViolet Coronagraph Spectrometer (UVCS; \citealt{Kohl1995}) on the Solar and Heliospheric Observatory (SOHO) satellite detected deviation from thermal equilibrium using spectral observations of the O{\footnotesize VI} 1032 and 1037\AA~doublet, or O$^{5+}$.  The oxygen was measured to flow at twice the proton speed while having a super mass-proportional temperature, $T_{O5+}/T_p >16$, and notable O$^{5+}$ temperature anisotropies, $T_{\perp}/T_{\parallel} > 10$, by 2R$_{\sun}$ \citep{Raymond1997, Cranmer2009_living_review}. Mg {\footnotesize X} 1219.71\AA, or Mg$^{9+}$, was also observed to be super heated, like oxygen. These observations suggest that non-thermal processes in coronal hole wind are more effective at heating O$^{5+}$ (and Mg$^{9+}$) compared to protons, and the heating is preferentially perpendicular to the local magnetic field direction. Several studies suggest that Alfv\'en cyclotron waves can resonantly transfer energy to the ions, leading to preferentially perpendicular heating as they dissipate that reconcile the heavy ion observations \citep{Hu1999, Hollweg2002, Kasper2008}, with recent near-Sun observations identifying their presence in the solar wind reinforcing this theory \citep{Liu2023, Bowen2024, Bowen2024b}. However, the process by which the waves are generated and how dissipation occurs remains difficult to pin down \citep{Marsch_Tu2001, Hollweg2002, Isenberg2007, Isenberg2009, Cranmer2003, Hellinger2005, Cranmer2019, Howes2024}. 

Other mechanisms have been proposed. Stochastic heating via dissipation of low-frequency turbulence could also explain the observed heavy ion temperatures \citep{Chandran2010, Chandran2013}. The so-called ``helicity barrier'' offers an avenue for ion cyclotron heating, in line with theory and observations, that may unify wave and turbulence theories addressing ion heating \citep{Meyrand2021, Squire2022NatAs, Zhang2025, Panchal2025}. Most recently, several studies explore the role of kinetic Alfv\'en waves in ion heating and acceleration in the $1-10R_{\Sun}$ regime that would also be compatible with coronal observations \citep{Ayaz2024, Ayaz2025, Ayaz2025b}.

The non-thermal characteristics extend and continue to evolve in the heliosphere. Several studies have found that the differential streaming of helium can be as large as the Alfv\'en speed (V$_A$), as observed with Helios 1 \& 2 between 0.3 and 1au \citep{Marsch1982, Durovcov2019}, while continuing to decrease beyond 1au as observed with Ulysses \citep{Neugebauer1996, Reisenfeld2001}.  Additionally, the scalar temperature of helium is observed to be higher than protons, to varying degrees, often exceeding mass proportionality i.e. becoming super-mass proportional \citep{Kasper2008, Maruca2013}. 

Most recently, observations from Parker Solar Probe (Parker; \citealt{Fox2016}) have provided new insight into the kinetics of helium below 0.3au \citep{Mostafavi2022}. Studies find that larger differential speeds between He$^{2+}$ and protons ($V_{He2+,p} = V_{He2+} - V_{p}$) are correlated to higher temperature ratios, T$_{He2+}$/T$_p$, at these distances \citep{Peng2024}. The observations in the study suggested a preferential heating zone for He$^{2+}$ that is below 0.16 au (or $\sim$ 34R$_{\sun}$).  This is in line with other studies that predict that the majority of preferential heating occurs below the Alfv\'en surface \citep{Kasper2017,Kasper_Klein_2019} while that distance varies for heavier ions \citep{Holmes2024}. Studies have also found that differential flow of He$^{2+}$ reaches a maximum of $100–200$ km s$^{‑1}$, despite an increasing Alfv\'en speed closer to the Sun \citep{Wang2025}. 

Prior to the launch of Solar Orbiter \citep{Muller2020} in 2020, heavier ions ($Z>2$ where Z is the atomic number or number of protons) have only been examined between $1-5$au with the Solar Wind Ion Composition Spectrometer (SWICS) on board both the Advanced Composition Explorer (ACE) and Ulysses, the Charge Time-of-Flight (CTOF) solar wind ion sensor on SOHO \citep{Hefti1998, Janitzek2016}, and STEREO/PLASTIC \citep{Bochsler2010, Bochsler2010b}. From ACE/SWICS observations, \cite{Berger2011} examined the differential flow ($V_{i,p} = V_{i} - V_{p}$) of 44 heavy ions at L1, across mass ($m=4-56$ amu) and mass-per-charge ($m/q = 2-8$ amu/e), for a two week period in a predominately fast wind stream. The study finds that, on average, the minor ions stream at 0.55V$_{A}$, with no clear $m/q$ or $m$ dependence. The majority of differential streaming was positive with a few instances of negative differential streaming that the authors attribute to kinks in the magnetic field. In the same systematic manner, \cite{Tracy2015, Tracy2016} examine the temperature and thermal speed of 50 minor ions, organized by collisional age, observed at 1 au from 13 years of data. The collisional age is an in situ, single-point estimate of the cumulative effect of Coulomb collisions -- normalizing the collision frequency to the transit time of a plasma parcel -- that would act to thermalize the solar wind and erase non-thermal signatures \citep{Kasper2008, Alterman2018,  Martinovic2021}, including any super-mass proportional temperature observed in the corona. Near 1 au, the temperature of O$^{6+}$ ($m/q = 16/6 = 2.67$), and several other ions, compared to that of protons (T$_i$/T$_p$) show a convergence towards unity at the highest collisional ages. However, the temperature ratio did not strictly decrease, suggesting more complex behavior across solar wind speed and its radial evolution.  

In addition to collisions, wave-particle interactions may explain the simultaneous decrease in differential flow and increase in temperature of the He$^{2+}$ population. For example, two-dimensional hybrid simulations of collisionless plasma show that wave-particle scattering, via alpha-proton parallel magnetosonic and Alfv\'en instabilities, reduces the relative flow between helium and protons, heats the helium more strongly than the protons, and increases $T_{\perp}/T_{\parallel}$ of the helium, in line with the radial profiles seen in He$^{2+}$ observations \citep{Gary2000_instabilities_diff_speed}. In addition, the theoretical framework of stochastic heating from \cite{Chandran2013} also finds an inverse relationship between differential flow and the perpendicular temperature ratio of He$^{2+}$ to protons, matching well with observations from the \emph{Wind} spacecraft \citep{Kasper2013}. However, it is unknown if heavier ions behave in the same manner and can be described the same way by these processes.

Further evaluation of several theories and models require extended knowledge of minor ion temperatures and speeds at several stages of solar wind outflow. As such, this work examines the radial evolution of O$^{6+}$ drift speed and temperature between $0.3-1$ au to analyze its behavior in the solar wind. 

\begin{figure*}
	\centering
	
	\includegraphics[width=0.49\textwidth]{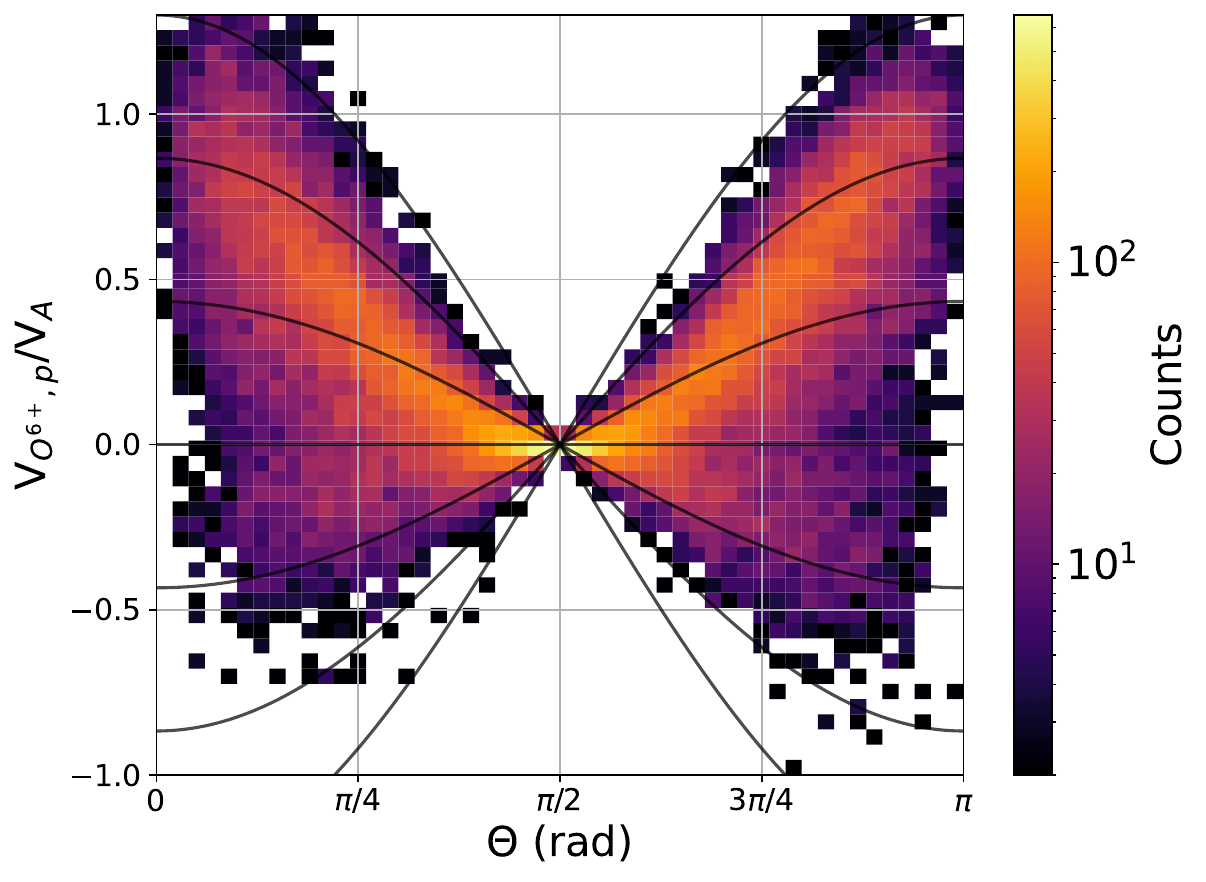}
	\includegraphics[width=0.49\textwidth]{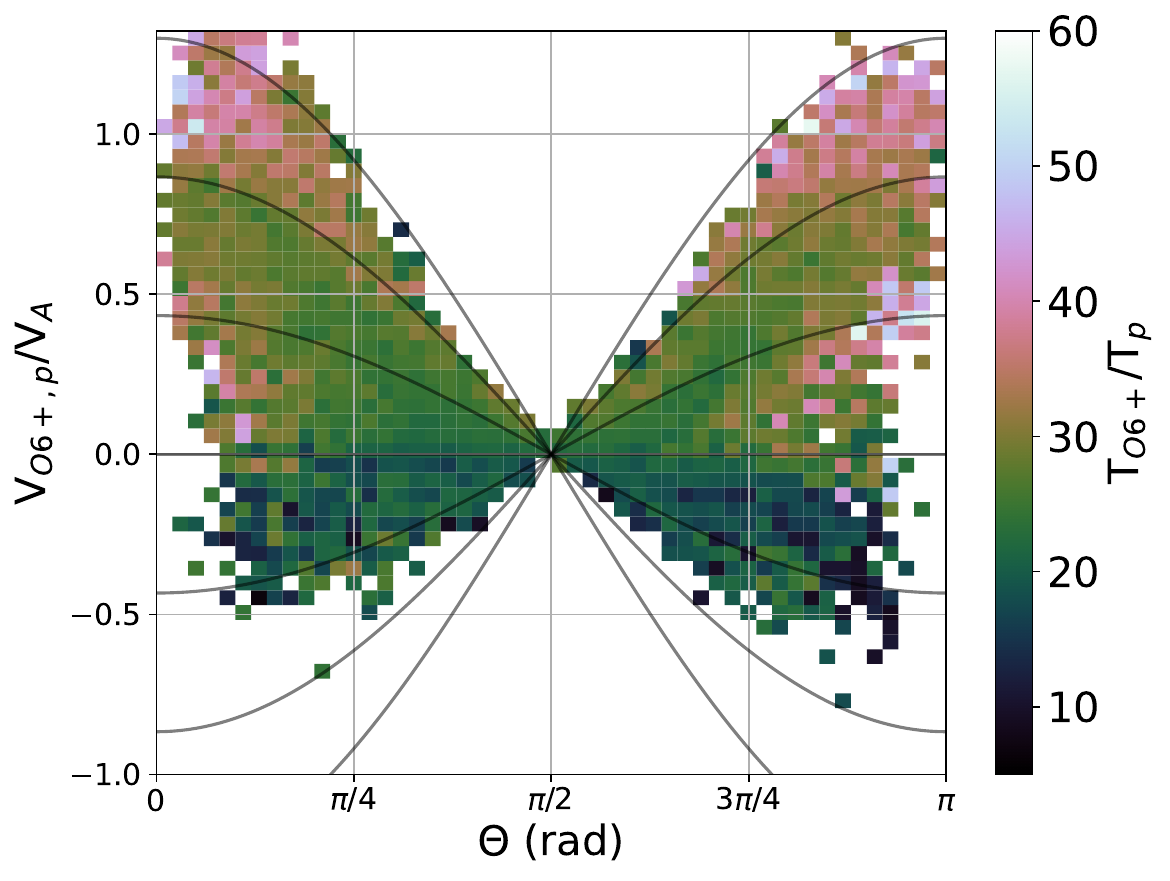}
	
	\caption{2D histograms of differential streaming of O$^{6+}$ normalized to the local Alfv\'en speed versus angle between the radial direction and the local magnetic field, which we refer to as a ``Moth plot''. The left panel shows the distribution of all measurements taken while the right panel shows the average temperature ratios (T$_{O6+}$/T$_p$) within the parameters space for bins that contain at least 5 points. We overplot (solid lines) Eq.~(\ref{diff_pred}) for different $\Delta U/V_A$ for reference with amplitudes of $\pm$1.3, 0.87, 0.43, 0.}
	\label{fig:all_differential_streaming}
\end{figure*}

\section{Observations} \label{sec:observations}
 The study utilizes observations taken by the Heavy Ion Sensor (HIS; \citealt{Livi2023}), part of the Solar Wind Analyzer (SWA; \citealt{Owen2020}) suite on Solar Orbiter, across a 1 year and 4 month period during the ascending phase of solar cycle 25. To examine the radial evolution of differential streaming and temperature together, we use O$^{6+}$ speed and temperature from V02 of the HIS Level 3 dataset (released December 2024\footnote{\url{https://www.cosmos.esa.int/documents/3689933/11863906/SWA-HIS_usage_notes_L3_V02.pdf}}). In this version, the HIS dataset includes 10-minute cadence observations of oxygen and carbon ion ratios, iron charge state distribution, as well as the Fe/O abundances ratio. This dataset also includes O$^{6+}$ bulk and thermal speeds, with more heavy ion kinematics planned for future releases. This study therefore focuses on these properties of O$^{6+}$.

The study uses measurements of the magnetic field and proton speed, temperature, and densities taken by the magnetometer on board (MAG; \cite{Horbury2020}) and Proton Alpha Sensor (PAS) of SWA, respectively. To compute the different quantities, the measurements from PAS and MAG have been averaged down to match the 10 minute HIS resolution.

The period examined spans from 1 January 2022 to 30 April 2023.  The trajectory of Solar Orbiter during this period is shown in Appendix Figure \ref{fig:context_orbits}. The period includes three perihelia, with closest approaches on 26 March 2022, 12 October 2022, and 10 April 2023. To ensure the majority of the analysis included only non-transient solar wind, we removed coronal mass ejections (CMEs) as identified in the Helio4Cast catalog \citep{Mostl2017, Mostl2020}\footnote{\url{https://helioforecast.space/icmecat}}. After the removal of CME periods, the number of 10 minute intervals considered in the study is $N=40\,613$.  

We compute differential streaming normalized to the Alfv\'en speed (V$_{O6+,p}$/V$_A$) using the magnitude of the proton speed in the spacecraft frame and of the O$^{6+}$ speed, where we assume the bulk speed is dominated by the radial component. We use the spacecraft frame, where -X component is aligned with the radial component (R) in the RTN system, such that the difference between O$^{6+}$ and protons eliminates the spacecraft speed.  The Alfv\'en speed is computed as V$_A= \frac{B}{\sqrt{\mu_{0}m_p n_p}}$ where $B$ is the magnetic field strength, $\mu_{0}$ is the permeability of free space, $m_p$ and $n_p$ are the mass and number density of protons.

We now develop a prediction for the dependence of the relative drift on the angle $\Theta$ between the radial direction and the local magnetic field. We assume that the relative drift of heavy ions is aligned with the magnetic field. Moreover, we assume that any dependence of the relative drift on $\Theta$ is due to radial variations of the magnetic field, e.g., in the form of localized kinks or large-scale variations. Therefore, we set

\begin{equation}\label{diff_flow}
\vec V_{O6+}=\vec V_p\pm \Delta U\,\hat{\vec b},
\end{equation}
where $\vec V_{O6+}$ is the bulk velocity of O$^{6+}$, $\vec V_{p}$ is the bulk velocity of the protons, $\Delta U$ is the magnitude of the  relative drift between O$^{6+}$ and protons, and $\hat{\vec b}=\vec B/B$ is the unit vector of the magnetic field. Assuming that heavy ions are preferentially accelerated in the corona, we choose the upper sign in Eq.~(\ref{diff_flow}) for measurements in a sector with anti-Sunward magnetic field connectivity, and the lower sign for measurements in a sector with Sunward magnetic field connectivity. As we measure the radial components of the bulk velocities $V_{O6+,r}$ and $V_{p,r}$, we take the scalar product of Eq.~(\ref{diff_flow}) with the unit vector $\hat{\vec r}$ in the radial direction and find

\begin{equation}\label{diff_pred}
\frac{V_{O6+,p}}{V_A}=\frac{V_{O6+,r}-V_{p,r}}{V_A}=\pm \frac{\Delta U}{V_A}\,\cos\Theta,
\end{equation}

where $\cos\Theta=\hat{\vec r}\cdot \hat{\vec b}$. Our measurement of $V_{O6+,p}/V_A$ thus depends on the relative drift $\Delta U/V_A$ in the field-aligned reference frame and the angle $\Theta$. For $0<\Theta<\pi/2$, we find $V_{O6+,p}/V_A>0$ in sectors with anti-Sunward magnetic field connectivity and $V_{O6+,p}/V_A<0$ in sectors with Sunward magnetic field connectivity. For $\pi/2<\Theta<\pi$, we find $V_{O6+,p}/V_A>0$ in sectors with Sunward magnetic field connectivity and $V_{O6+,p}/V_A<0$ in sectors with anti-Sunward magnetic field connectivity. Therefore, we predict that all measurements with $V_{O6+,p}/V_A<0$ represent times in which the magnetic field is locally directed opposite to the direction associated with the global sector (e.g., through a local kink in the field that changes the sign of $B_r$). 

We compute the O$^{6+}$ scalar temperature from the thermal velocity, as: 
\begin{equation}
\begin{split}
    v_{O6+, th} = \sqrt{\frac{2k_BT_{O6+}}{m_{O6+}}} \\ 
    T_{O6+} = \frac{mv_{O6+,th}^2}{2k_B}
\end{split}
\end{equation}
where k$_B$ is the Boltzmann constant, m$_{O6+}$ and $v_{O6+, th}$ is the O$^{6+}$ mass and thermal speed.

\begin{figure*}
\centering

\includegraphics[width=0.48\textwidth]{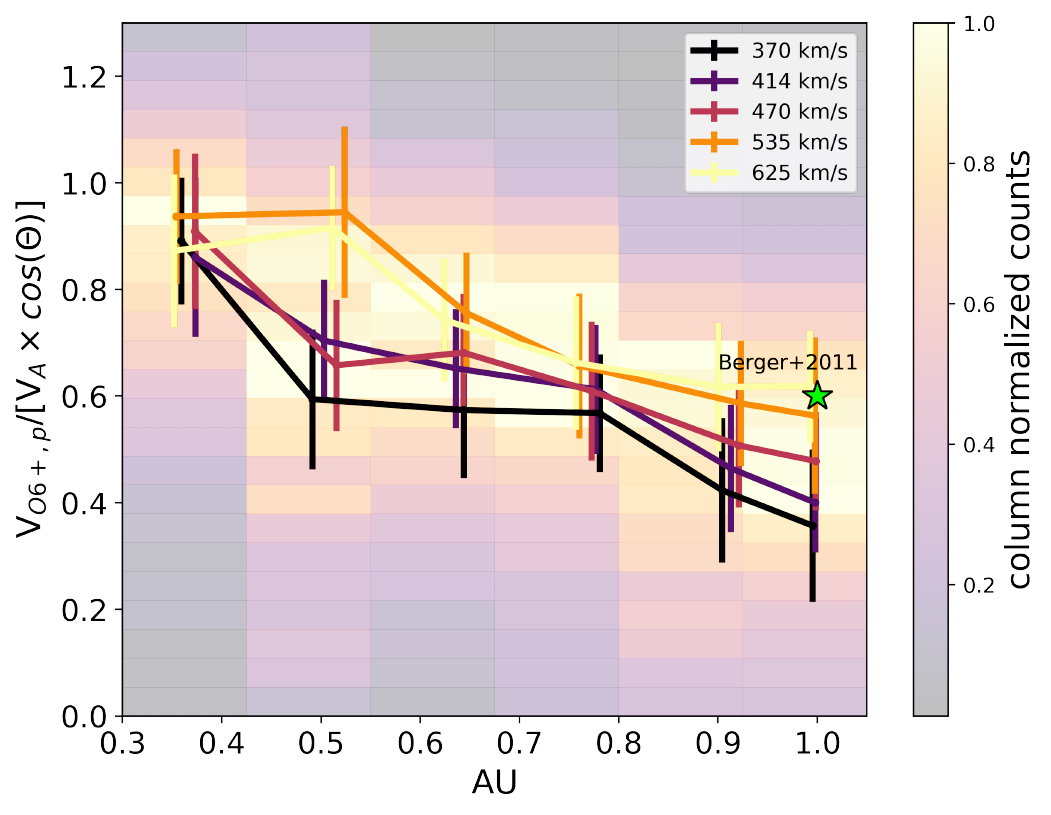}
\includegraphics[width=0.48\textwidth]{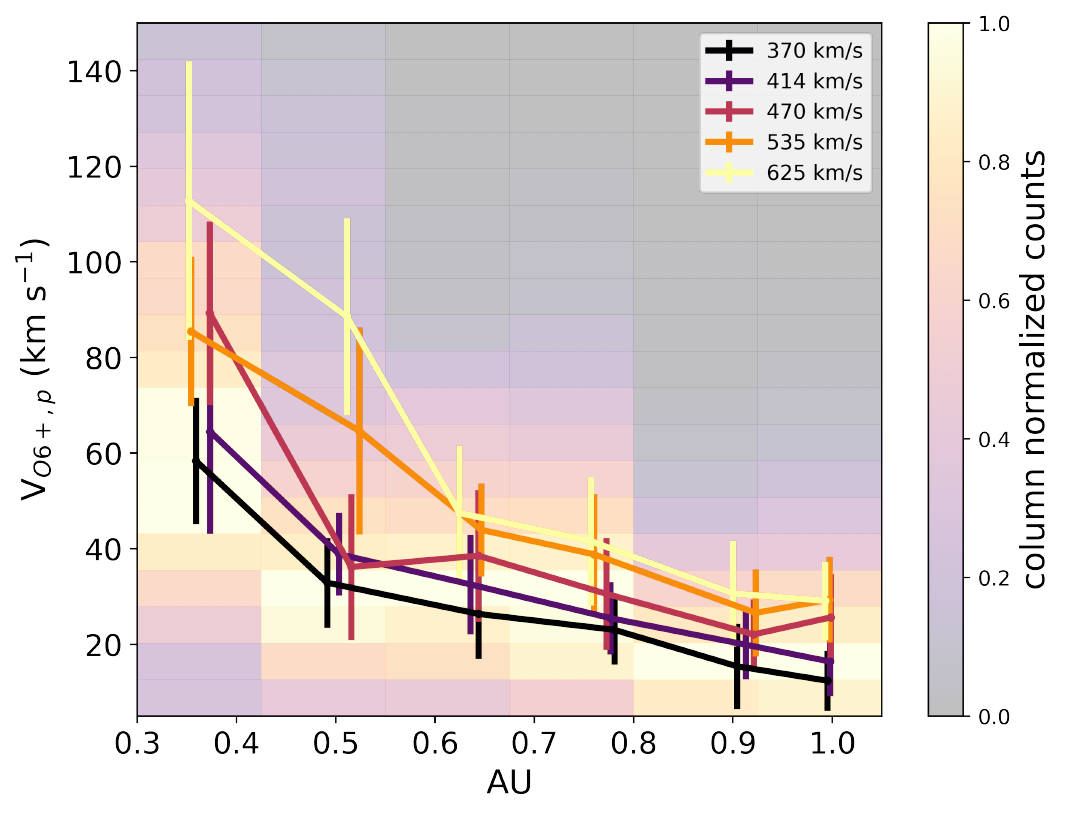}
\includegraphics[width=0.48\textwidth]{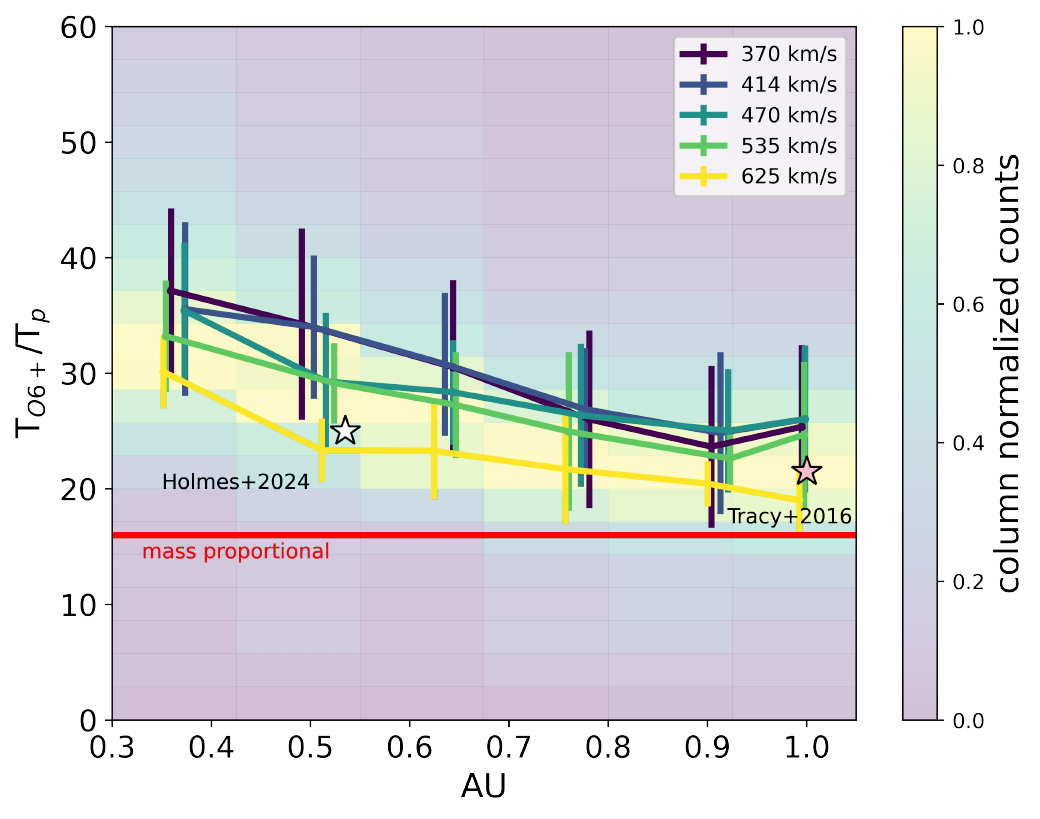}
\includegraphics[width=0.48\textwidth]{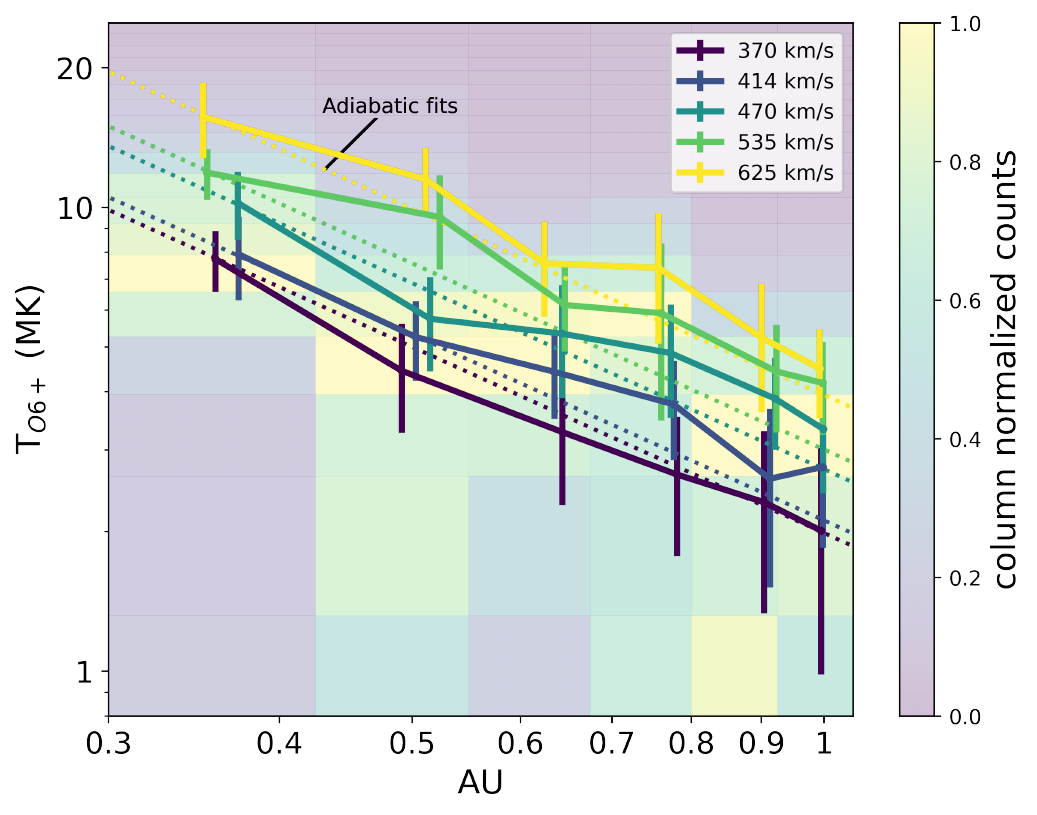}

\caption{2D histograms and radial profiles of O$^{6+}$ (top left) differential flow normalized to the local Alfv\'en speed $\times$ $\cos{\Theta}$, (top right) differential streaming, (bottom left) scalar temperature ratio with protons, and (bottom right) temperature, all plotted against heliocentric distance from the Sun in au. The bottom right plot has a log-log scale while the others are in linear scale. The background histograms are a distributions of counts in this parameter space that are column normalized. The radial profiles have been binned by proton speed, where the mean is reported in the legend, see Section \ref{sec:results} for bin ranges. In the top and bottom left plots, we include observations of O$^{6+}$ from \cite{Berger2011} and \cite{Tracy2016} as well as a prediction from \cite{Holmes2024}.}
\label{fig:radial_profiles}
\end{figure*}

\section{Results} \label{sec:results}

Figure \ref{fig:all_differential_streaming} presents 2D histograms, referred to as ``Moth plots'', that show the distribution of counts (left) and the corresponding mean temperature ratio (T$_{O6+}$/T$_p$, right) as a function of differential streaming (V$_{O6+,p}$/V$_A$) and angle between the radial direction and the magnetic field ($\Theta$) in radians. We overplot Eq.~(\ref{diff_pred}) for different $\Delta U/V_A$ for reference with amplitudes of 1.3, 0.87, and 0.43. The plot on the left incorporates all points measured and the right plot only includes bins that contain at least 5 counts to compute the mean of the temperature ratio.

We find that in contrast to \cite{Berger2011} Figure 3, both plots of Figure \ref{fig:all_differential_streaming} show the parameter space is filled-in simply by virtue of including a longer timeframe of observations, with more solar wind conditions. We also note that the figure incorporates differential streaming of all solar wind streams measured between $0.3-1$ au, with observations including different stages of evolution that collapses time and space. Despite this, we see organized behavior in the differential streaming when viewed together.  The largest differential flows are observed during field-aligned periods that decrease as the magnetic field is perpendicular to the flow, consistent with our prediction in Eq.~(\ref{diff_pred}). As in \cite{Berger2011}, we also find periods of negative differential streaming in O$^{6+}$, albeit not symmetrically distributed between positive and negative differential flows. This is discussed further in Section \ref{sec:negativediffstream}.

There is a significant population of solar wind with differential streaming speeds that are larger than the local Alfv\'en speed, requiring $V_{O6+,p}$ 30\% larger than $V_A$. Through a comparison with the O$^{6+}$ to proton temperature ratio, in the right panel of Figure \ref{fig:all_differential_streaming}, we find that larger differential speeds correspond to the largest average temperature ratios. The temperature ratio decreases with lower differential speed, with the lowest ratios within periods of negative differential streaming. To disentangle how the O$^{6+}$ temperature and relative flow change with distance from the Sun and solar wind speed within the Moth plot parameter space, we also include Appendix Figures \ref{fig:differential_streaming_versus_distance} and \ref{fig:differential_streaming_versus_distance_temp} in Appendix Section \ref{appsec:moth_plots} that sub-categorizes both panels from Figure \ref{fig:all_differential_streaming}. 

We further examine the radial evolution of O$^{6+}$ differential flow and temperature in Figure \ref{fig:radial_profiles}. The figure includes four panels showing the absolute values of differential streaming normalized to the local Alfv\'en speed $\times$ $\cos{\Theta}$ (top left) and just the speed difference (top right) as well as the temperature ratio (T$_{O6+}$/T$_p$) and O$^{6+}$ temperature (bottom row), all with increasing distance from the Sun. In the top left plot, we divide the relative flow by $\cos{\Theta}$ and only include $\Theta< 1$ and $>2.2$ (rad) to avoid dividing near 0 and only include the positive differential streaming periods. The purpose is to remove variability of differential streaming due to fluctuations that act to reduce the projected relative flow and to only consider the background speed difference (later discussed in Section \ref{sec:negativediffstream}). For each case, the solar wind is binned by proton speed (5 speed quantiles) and distance from the Sun (7 equally spaced bins) that range as: 349-389, 390-440, 441-500, 501-574, and 575-900 km s$^{-1}$ and 0.3-0.43, 0.44-0.57, 0.58-0.7, 0.71-0.83, 0.84-0.97, 0.98-1.1 au. The background of each plot shows the column normalized total counts within the parameter space. The vertical bars for each is computed as the standard deviation of the mean.

Generally, all the parameters plotted in Figure \ref{fig:radial_profiles} decrease with distance from the Sun. As has been observed with He$^{2+}$ in the past at 0.3au and beyond \citep{Marsch1982, Mostafavi2022}, O$^{6+}$ is also observed to decrease in its differential streaming and to have a comparable speed differences (V$_{O6+,p}$, as in top right panel). In this analysis, some speed dependence arises, although the overall decrease in differential streaming behavior is similar across all solar wind measured. At 1au, our results show that the range of mean differential flow (top left) across all wind speeds span between $0.4-0.8$V$_A$. At 1au, \cite{Berger2011} finds an average differential flow of heavy ions to be $0.55$V$_{A}$ with O$^{6+}$ being slightly larger than the average at $0.6$V$_{A}$, as indicated by the green star in the top left panel of the figure. Overall, the differential steaming from the present study (for the fastest speed bin, yellow curve) and ACE/SWICS from \cite{Berger2011} (mainly for a fast wind stream) at 1au coincide well. 

For the temperature (bottom row of Figure \ref{fig:radial_profiles}), we find that the temperature ratio decreases while remaining super-mass proportional, (T$_{O6+}$/T$_p > 16$), out to 1au. Additionally, we find that the O$^{6+}$ population on average cools adiabatically between $0.3-1$au, as indicated by the $T\propto r^{-4/3}$ fitted profiles (represented as dashed lines for each speed bin), assuming an adiabatically expanding ideal gas \citep{Schwartz1983}. The profiles use an initial temperature from 0.3au as the first value of each curve and are extrapolated to 1au. Together, these observations suggest strong preferential heating of O$^{6+}$ has finished well below 0.3au and, directly shows, that no significant local heating occurs beyond this distance. Furthermore, we compare observations to previous studies, \cite{Tracy2016} as indicated by the pink star and a prediction from \cite{Holmes2024} shown in the white star of the bottom left panel, that is discussed in Section \ref{sec:preferential heating}.

 \begin{figure*}
\centering
\includegraphics[width=0.47\textwidth]{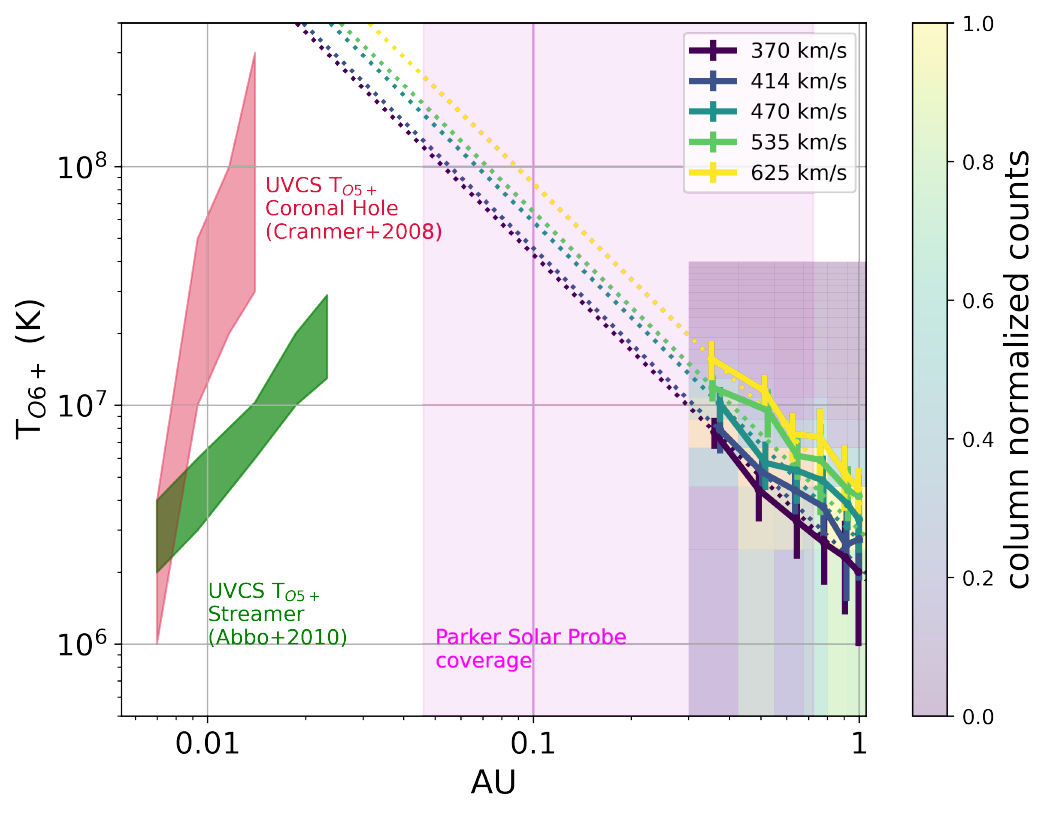}
\includegraphics[width=0.47\textwidth]{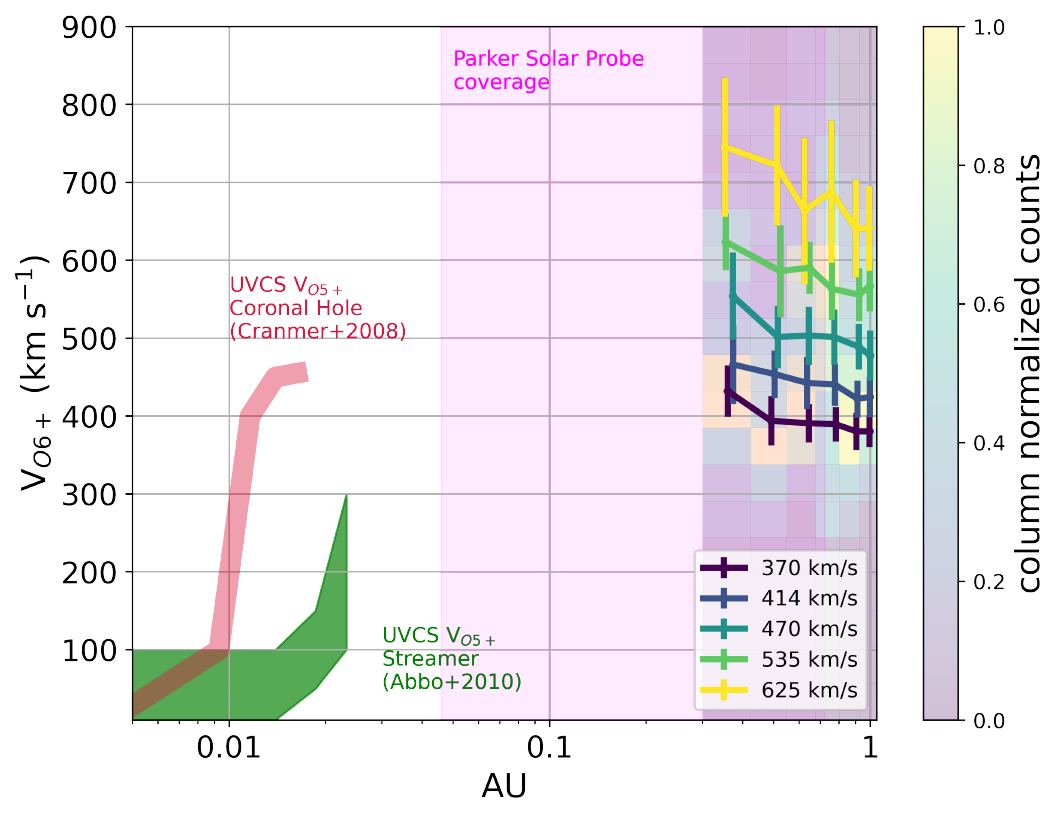}
\caption{High charge states of oxygen in the inner heliosphere: O$^{6+}$ temperature and speed versus heliocentric distance to the Sun between $0-1$au. The plots include observations from UVCS of O$^{5+}$ in a polar coronal hole from \cite{Cranmer2008} and a streamer from \cite{Abbo2010, Abbo2016}, whose ranges are indicated by the red and green regions respectively, extending between approximately from the solar surface to $0.0186$au ($1-4R_{\Sun}$). We also include the coverage of Parker Solar Probe between $\sim9.86-155R_{\sun}$.}
\label{fig:compare_UVCS}
\end{figure*}

\section{Discussion} \label{sec:discussion}

\subsection{O$^{6+}$ temperature and relative flow}
Generally, we find that the average differential flow between O$^{6+}$ and protons is reduced with radial distance while its radial scalar temperature profile indicates no strong heating, as shown in Figure \ref{fig:radial_profiles}. This behavior deviates from what is observed in He$^{2+}$ radial studies, where T$_{He2+}$ shows non-adiabatic cooling similar to protons, while reducing its differential flow \citep{Durovcov2019, Peng2024}. In our analysis, we note that an adiabatic temperature profile, ($1/r^{4/3}$), may not be strictly true for a turbulent, collisionless plasma with strong temperature anisotropies such as the solar wind which can drive instabilities and can depart from an ideal gas \citep{Matteini2007, Matteini2012SSRv}. Statistically, temperature anisotropies ($T_{\perp}/T_{\parallel}$) are observed for protons and alpha particles and increase for both with solar wind speed \citep{Marsch1982b}. Several studies have examined the role of instabilities as the driver of proton and helium temperature and differential flow behavior \citep{Gary2000_instabilities_diff_speed, Maruca2012, Chandran2013, Verscharen2013, McManus2024, Martinovic2024}. In linear Vlasov theory, proton and alpha temperature anisotropies can both be constrained by various kinetic instabilities: cyclotron, mirror, and parallel and oblique firehose. These instabilities impose limits to their temperature anisotropy ratios that inhibit unbounded growth or decay. Observations of minor ions indicate strong temperature anisotropies in the corona which may extend to the solar wind, driving instabilities that modulate its observed scalar temperature. Therefore, in addition to the scalar temperature examined in the present study, it is crucial to inspect O$^{6+}$ temperature anisotropy in more detail to account for anisotropic heating that is not directly observable in the present study.

\subsection{O$^{6+}$ Zone of Preferential Heating} \label{sec:preferential heating}
Comparison between the in situ speed and temperature profiles against those observed at the Sun provide some bounds to the zone of preferential oxygen heating. The radial temperature profiles and adiabatic temperature fits of O$^{6+}$ suggest the majority of heating it experiences occurs below 0.3au. Figure \ref{fig:radial_profiles}, bottom-left panel, that shows the temperature ratio (T$_{O6+}$/T$_p$) includes the temperature ratio predicted by \citet{Holmes2024} for the outer radial boundary of the zone of preferential heating for O$^{6+}$. The outer radial boundary represents the location where preferential heating of minor ions is active. The value predicted by \citet{Holmes2024}, represented by a white star, is on the order of 25 and falls within the range of average values computed for the fastest wind (yellow curve). However, the value from \cite{Holmes2024} would not be compatible with excess local heating up to the predicted outer radial boundary of 115$R_{\Sun}$ (0.54au) given that the temperature shown in the bottom right panel is well described by adiabatic cooling starting at 65$R_{\Sun}$ (0.3au). 

Additionally, we compare our results to temperature observations at 1au from \cite{Tracy2016}. The study finds an average behavior of $\frac{T_i}{T_p}=\frac{4}{3}\frac{m_{i}}{m_{p}}$ for ion, $i$, as determined through a simple linear fit to heavy ion temperatures from 13 years of observations. Therefore, the temperature ratio for O$^{6+}$ should be on the order of 21.3 at the outer edge of our observation domain, as indicated by the pink star at 1au in the figure. Our results show the average temperature ratio and its variability across wind speeds falls within the predicted value of \cite{Tracy2016}. 

\subsection{Comparison to Remote Observations of Oxygen}
Figure \ref{fig:compare_UVCS} shows a comparison of the present analysis of O$^{6+}$ (taken during the ascending phase of the solar cycle) to O$^{5+}$ temperature and speed at the Sun observed in a polar coronal hole (red) with UVCS/SOHO between 1996 and 1997 \citep{Cranmer2008} and a equatorial streamer (green) in solar minimum \citep{Abbo2010,Abbo2016}.  We note that heavy ion properties in the slow solar wind do exhibit some solar cycle dependence \citep{Carpenter2024, Alterman2025a, Alterman2025b} therefore some variation between the remote observations of the streamer (solar min) being connected slow solar wind properties (ascending phase) in the heliosphere may be present. The figure also includes the present Parker coverage that spans $9.86R_{\Sun}$ to $155R_{\Sun}$ for reference. Within the remote observations, the temperature increases between $1\times10^6$ to $2\times10^8$ K while increasing in speed from 20 to 500 km s$^{-1}$ across a heliocentric distance of $1-3.5R_{\sun}$ in the coronal hole. At the outer edge of the field of view, the temperature ratio of the coronal hole would correspond to T$_{O5+}$/T$_p = 23.5$ \citep{Cranmer2020b}. In the streamer, the temperature increases from 2$\times10^6$ to $3\times10^7$ and its outflow speed increases more gradually, reaching just below 300 km s$^{-1}$ by $5R_{\sun}$. At the outer edge of the streamer measurements, the average T$_{O5+}$/T$_p = 21$. 

The extrapolations of the O$^{6+}$ adiabatic temperature fits back to the Sun show that the temperature, if simply adiabatically expanding and cooling, would exceed the measured values observed in the low corona. The fitted profiles (if valid) would suggest that the temperature of O$^{6+}$ would continue to rise, peak, and turn over just after leaving the field of view of UVCS at $3.5-5R_{\sun}$. Under the assumption of a nearly constant proton temperature in the corona, this would also be supported by the measured temperature ratio of T$_{O5+}$/T$_p = 23.5$ and 21 between $3.5$ and $5R_{\sun}$. This ratio is lower and would be required to increase to match observations at 0.3au which falls between $30-40$ as shown in Figure \ref{fig:radial_profiles}, bottom left panel. However, this is difficult to disentangle because there are no observations of heavy ion properties above a few solar radii from the solar surface or below $65R_{\sun}$.

In the speed profile shown in the right panel of Figure \ref{fig:compare_UVCS}, the edge of the UVCS field of view sees a plane of the sky outflow velocity of 500 km s$^{-1}$ in a coronal hole and up to 300 km s$^{-1}$ in a streamer for O$^{5+}$. The in situ O$^{6+}$ speed is measured to be between 400-800 km s$^{-1}$ near 0.3au, where the yellow curve in the figure at 700-800 km s$^{-1}$ would best coincide with polar coronal hole observations, while the dark purple curve would correspond to the streamer coronal observations. Together, these observations suggest that O$^{5+}$ and O$^{6+}$ experience acceleration beyond the UVCS field of view in order to match the speed observed in situ.

\subsection{Negative Differential Streaming Behavior During Alfv\'enic Fluctuations} \label{sec:negativediffstream}

\begin{figure*}
\centering

\includegraphics[width=0.28\textwidth]{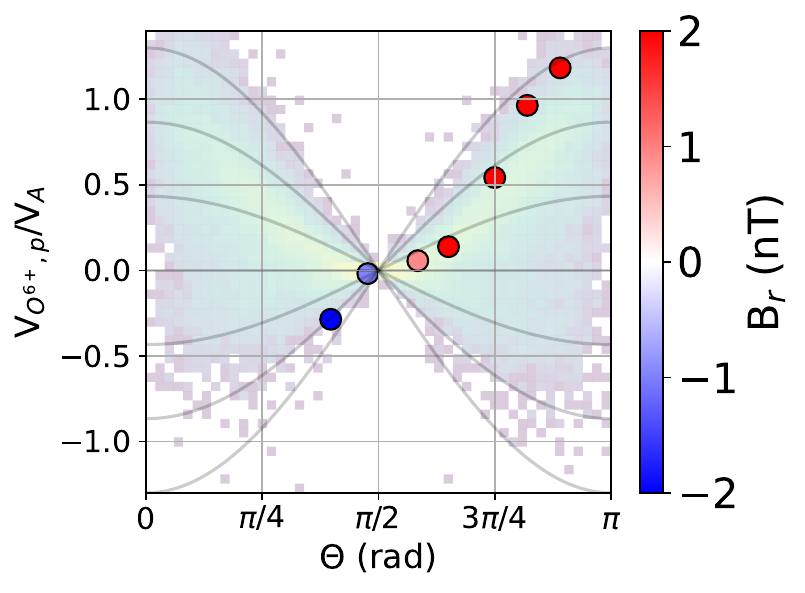}
\includegraphics[width=0.28\textwidth]{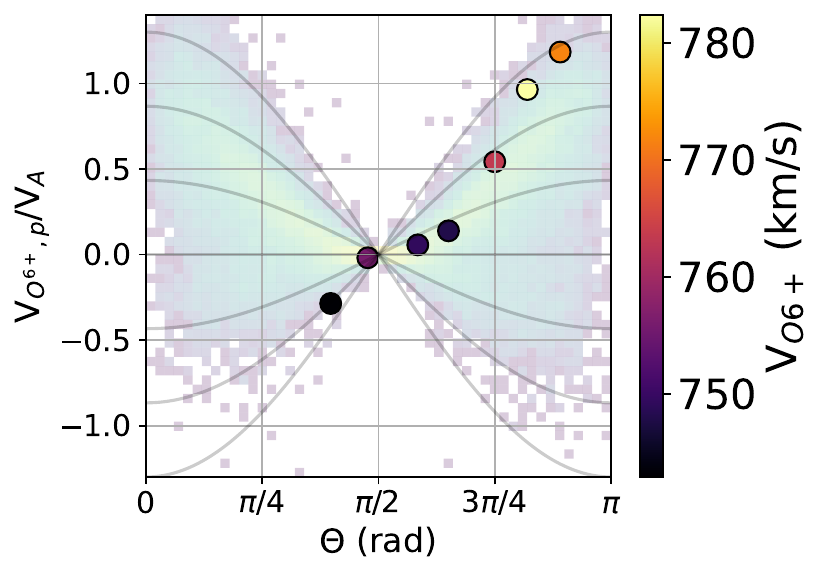}
\includegraphics[width=0.28\textwidth]{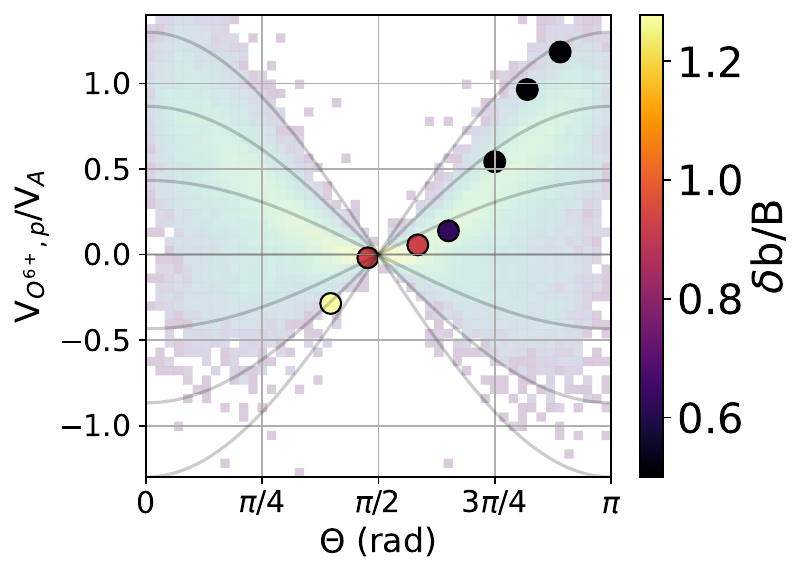}
\includegraphics[width=0.48\textwidth]{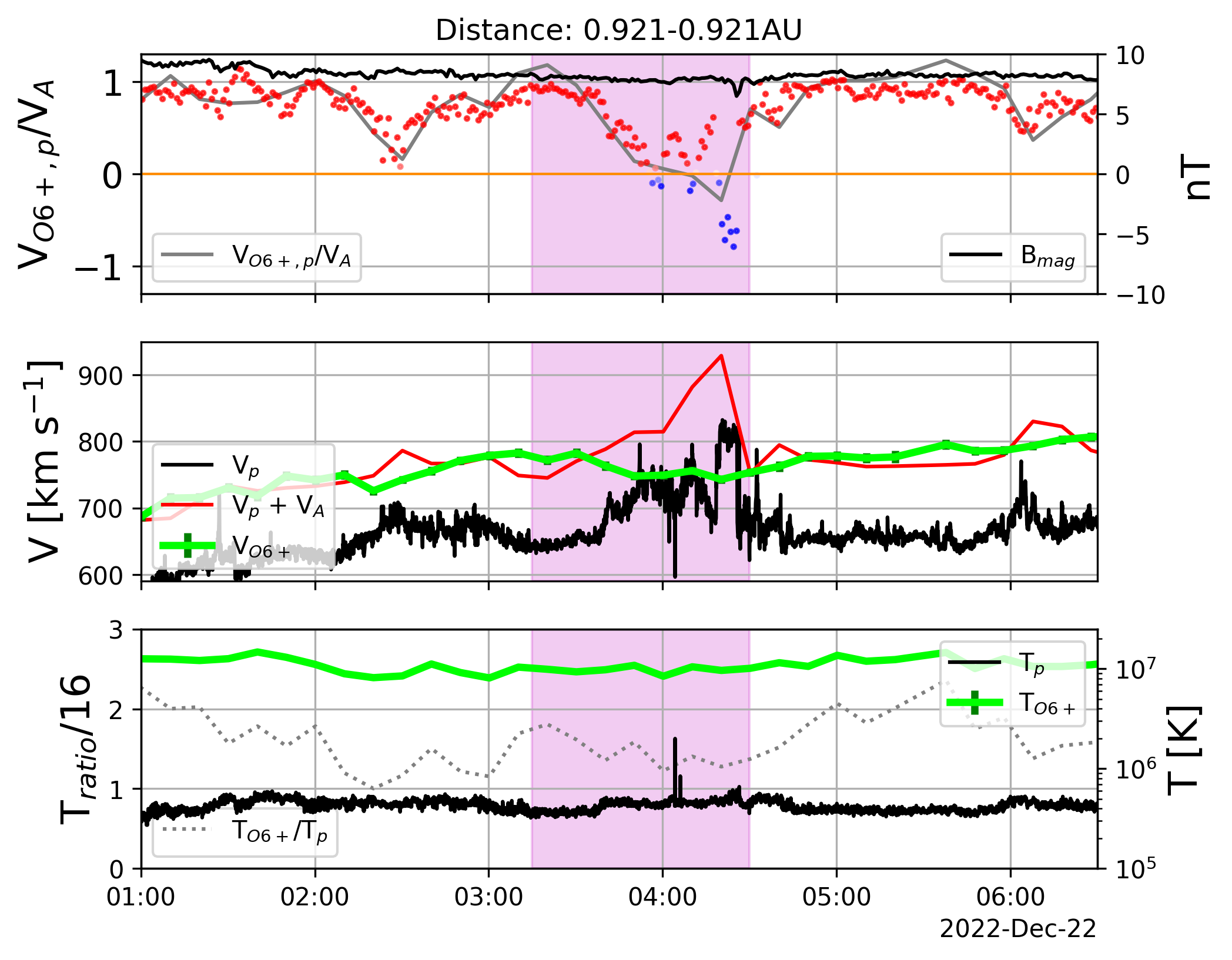}
\includegraphics[width=0.33\textwidth]{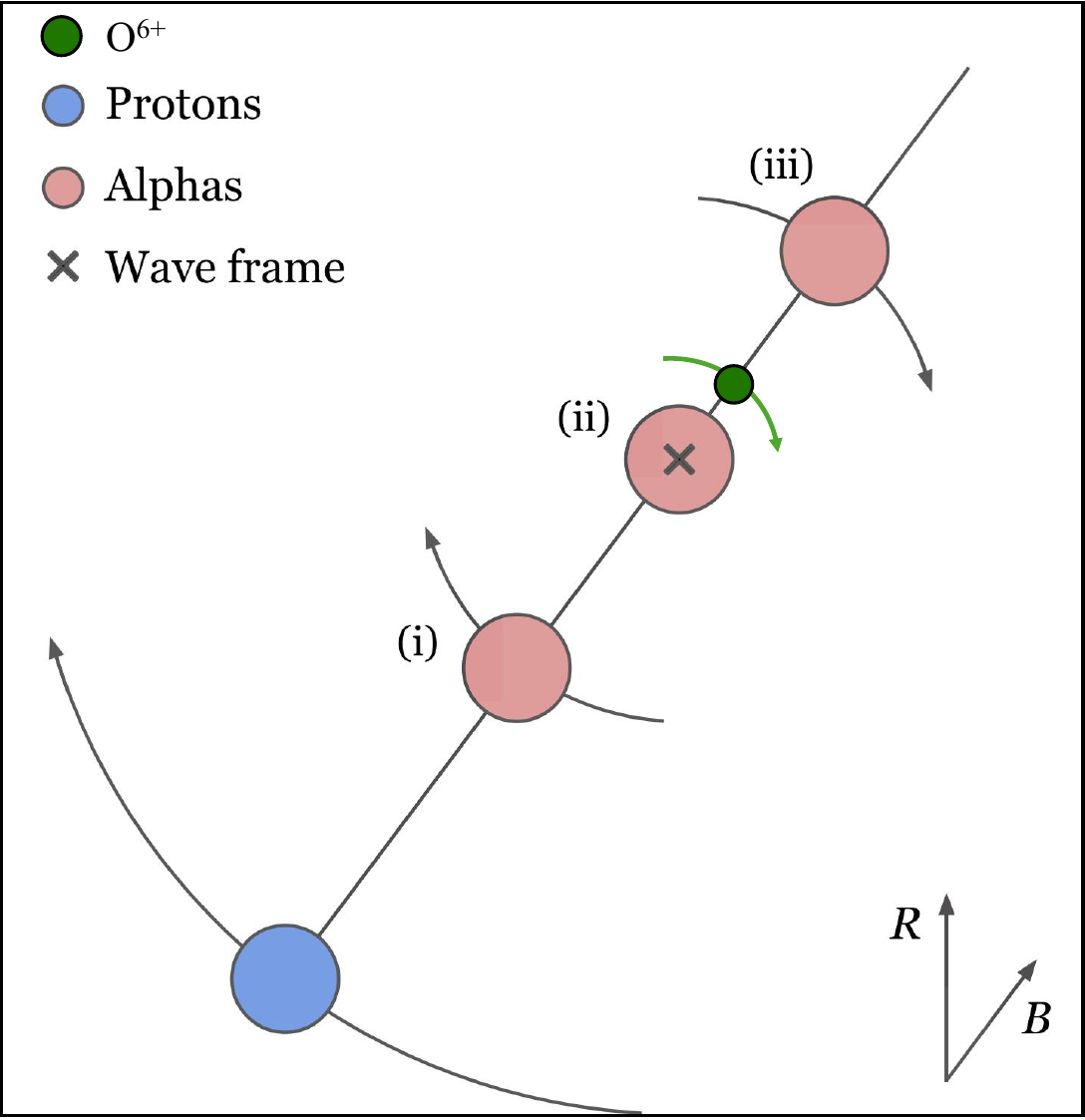}

\caption{Negative differential streaming periods due to an Alfv\'enic fluctuation. Top row shows three panels of 2D histograms with the radial component of the magnetic field, V$_{O6+}$, and $\delta B/B$ values across an individual Alfv\'enic fluctuation period highlighted in the shaded region shown in the timeseries below. The timeseries plots the top panel shows differential streaming in gray in association with the left vertical axis, and the radial magnetic field, B$_R$ (where red is for positive and blue is for negative polarity), and magnitude, B in black. The middle panel shows the proton (black) and O$^{6+}$ (green) speed magnitude along with the proton + Alfv\'en speed (red). The bottom panel shows the proton and O$^{6+}$ individual temperature (right vertical axis, black and green, respectively) along with their ratio divided by 16 (gray) associated with the left axis, to show mass proportionality. The cartoon figure on the bottom right is adapted from \cite{McManus2022} for the case in this example, see details in the text. }
\label{fig:ex1}
\end{figure*}

\begin{figure*}
\centering

\includegraphics[width=0.28\textwidth]{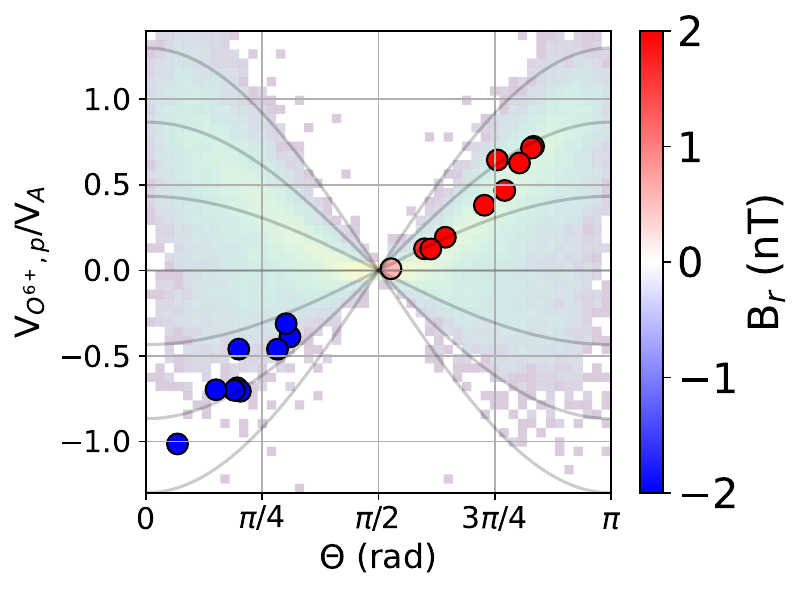}
\includegraphics[width=0.28\textwidth]{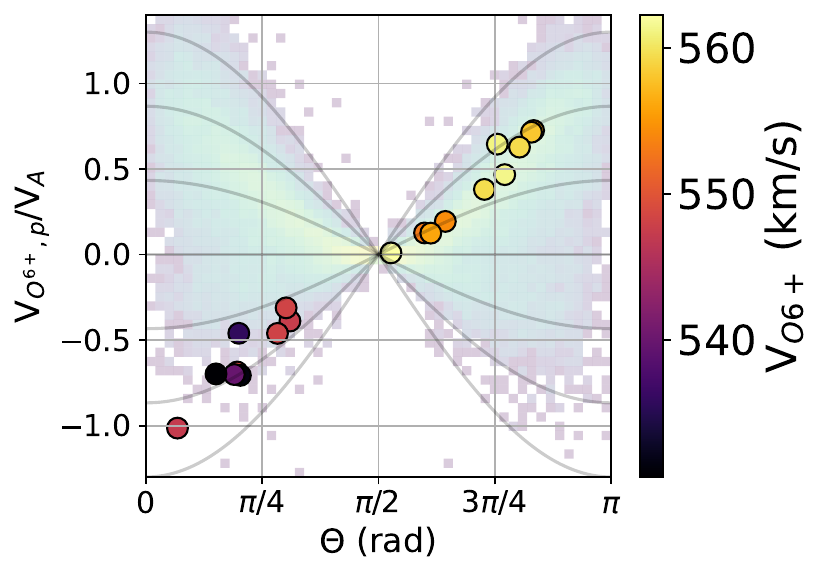}
\includegraphics[width=0.28\textwidth]{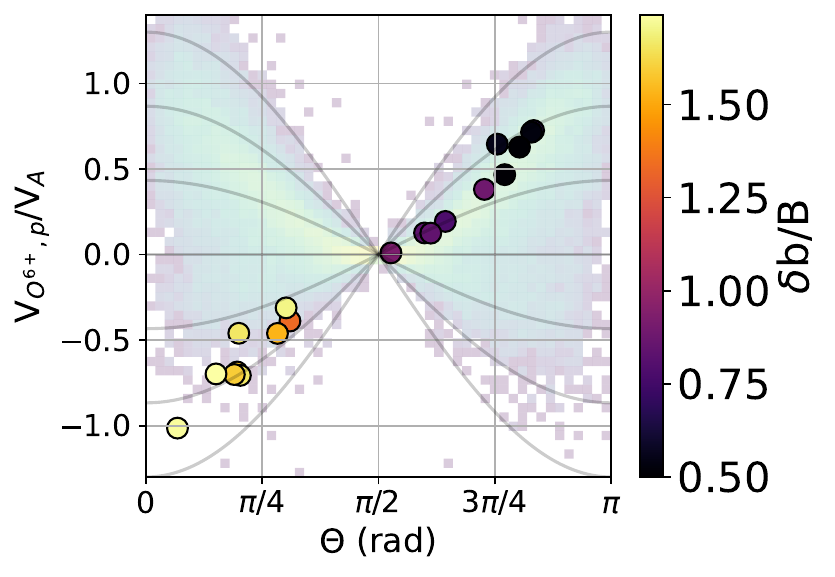}
\includegraphics[width=0.48\textwidth]{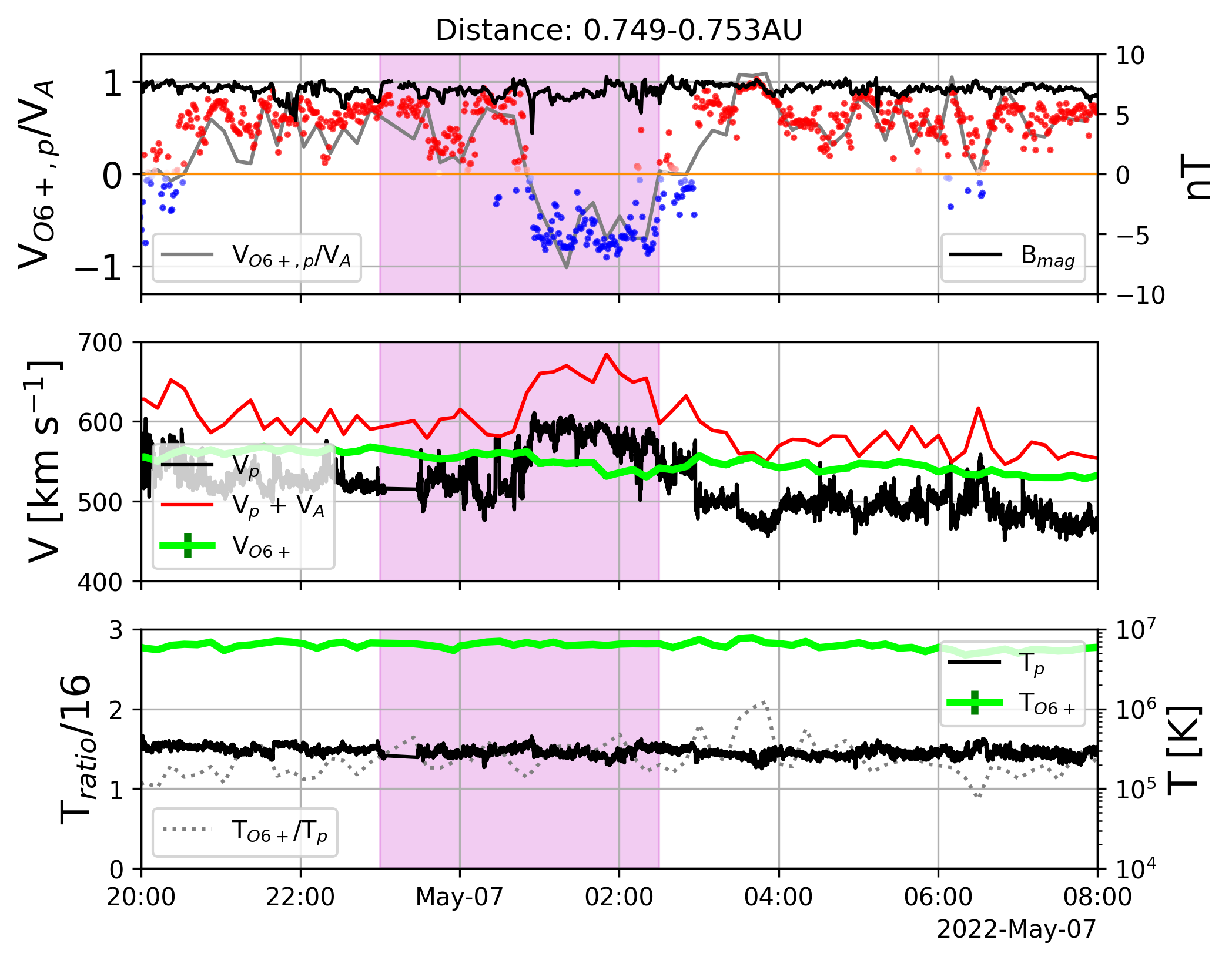}

\caption{Same as Figure \ref{fig:ex1} for a different case. }
\label{fig:ex2}
\end{figure*}

As observed in Figure \ref{fig:all_differential_streaming} and Appendix Figure \ref{fig:differential_streaming_versus_distance_temp}, there are case studies of significant negative differential flow between O$^{6+}$ and protons related to local reversals in the magnetic field, as predicted from Section \ref{sec:observations}. The magnetic reversals are Alfv\'enic fluctuations that show a correlated magnetic and velocity fluctuation that indicates high cross-helicity. Figure \ref{fig:ex1} and \ref{fig:ex2} shows examples during these periods that occurred on 22 December 2022 near 0.921au and 22 April 2023 between $0.384-0.39$au. The top row of panels shows the moth plot from Figure \ref{fig:all_differential_streaming} in the background, $\cos(\Theta)$ curves, and the measurements of differential streaming versus $\Theta$ across the shaded magenta period of the timeseries below. The magenta period encompasses the fluctuation period of interest. The colors in each top panel correspond to the colorbar showing the magnetic field polarity (red is positive and blue is negative, in top left), O$^{6+}$ speed in km s$^{-1}$ (top middle), and magnetic fluctuation normalized to the two-hour mean field magnitude ($\delta b/B$, top right). We compute $\delta b$ as $|\delta \textbf{b}| = \textbf{B}-<\textbf{B}>_t$, where $<\textbf{B}>_t$ is time-averaged over 2 hour interval, $t$. The timeseries (bottom left) shows the magnetic field magnitude (black) and B$_R$ (red/blue) along with V$_{O6+,p}$/V$_A$ (gray). These measurements were taking in a sector of anti-Sunward magnetic field connectivity. The middle panel contains the proton and O$_{6+}$ speed and V$_p$+V$_A$ (red) and the bottom panels are the individual temperature and ratio. Lastly, in Figure \ref{fig:ex2} the bottom right panel shows a modified image of a lever-arm picture from \cite{McManus2022} that predicts the velocity fluctuation behavior of He$^{2+}$ for three scenarios where the alphas flows slower (i), at the same speed (ii), and faster (iii) than the local wave speed, that would result as either an increase, no change, and decrease in the alpha speed, respectively (see also, \citealt{Matteini2014, Matteini2015, Horbury2018}). We modify the image to include O$^{6+}$ under the case of scenario (iii) in accordance to this example period. For the example fluctuation in Figure \ref{fig:ex2}, we find that the O$^{6+}$ behaves in accordance to the lever-arm scenario where with an initial differential flow faster than the local Alfv\'en speed, the O$^{6+}$ speed decreases, from 780 to 740 km s$^{-1}$ (top middle panel) as would be predicted by the cartoon picture in the bottom right and our simple model calculation in Section \ref{sec:observations}. 

However, for the second example shown in Figure \ref{fig:ex2}, in solar wind farther from the Sun, we find that despite the O$^{6+}$ differential flow speed being just below the Alfv\'en speed prior to the field rotation, the O$^{6+}$ speed decreases during the fluctuation while it is predicted to increase. For the period, the initial V$_{O6+,p}$/V$_{A}$ is shown in each panel of the top row (starting in the top-right portion of the plots) and observed to be lower than 1. For an O$^{6+}$ differential flow speed of less than 1, the lever-arm behavior would indicate  O$^{6+}$ speed would increase under scenario (i) from \cite{McManus2022}. However, as shown in the top middle panel, O$^{6+}$ speed decreases from 560 to 520 km s$^{-1}$. We interpret this as a case where the Alfv\'en speed calculated is larger than the actual wave speed, as has been previously discussed and observed in the solar wind \citep{Goldstein1995, Newgebauer1996_wave}. Depending on the difference, a lower wave speed (V$_w$) versus V$_A$ would increase the V$_{O6+,p}$/V$_w$ ratio computed prior to the fluctuation to a value larger than 1, indicating scenario (iii) rather than (i) from \cite{McManus2022} that would be compatible with observations. To explore this idea, we plot the magnetic field and velocity components, radial and perpendicular (projection of tangential and normal components, as $B_{\perp} = \sqrt{B_{T}^2 + B_{N}^2}$ and  $V_{\perp} = \sqrt{V_{T}^2 + V_{N}^2}$), illustrated in Appendix Figure \ref{fig:radial_perp_B_V}. We find the magnetic field components trace a circle (or sphere in 3D) across the fluctuation as would be the case for spherically polarized Alfv\'en waves. The velocity fluctuation indicates that the motion of the components have more curvature in the perpendicular components compared to the magnetic field.  However the clumps of data trace at least two coherent spheres in 3D. An inner circle is a curve fitted to the velocity covering the inner part of the $V_R$ and $V_{\perp}$ with a radius of 43.4 km s$^{-1}$, while V$_A=62$ km s$^{-1}$. The substructure that traces different spherical shells across the fluctuation could potentially indicate a distinct wave and Alfv\'en speed, and ultimately affect our lever-arm interpretation. Further analysis of individual cases across the different distances and wind speeds warrants its own detailed future study.

\section{Conclusions and Outlook} \label{sec:conclusions}
We examine the radial evolution of the O$^{6+}$ speed and temperature between $0.3-1$au within a 1.3 year timeframe of data collected by HIS on Solar Orbiter. The main conclusions of the study include that on average:

\begin{enumerate}
    \item Across all wind speed, differential flow, (V$_{O6+,p}$ and V$_{O6+,p}$/V$_A$) decrease with heliocentric distance from the Sun and is along the magnetic field direction
    \item The temperature ratio (T$_{O6+}$/T$_p$) decreases with distance but remains super-mass proportional ($>16$)
    \item T$_{O6+}$ cools adiabatically with increasing heliocentric distance across all wind speeds, suggesting no significant heating is experienced by O$^{6+}$ beyond 0.3au
    \item Negative differential streaming, where O$^{6+}$ tends to travel slower than the protons, increases with distance from the Sun and occurs during momentary kinks in the magnetic field, that are commonly associated with large amplitude Alfv\'enic fluctuations
    \item O$^{6+}$ can differentially flow faster than the local Alfv\'en speed and this is predominately observed to happen near perihelion periods, inside 0.57au
\end{enumerate}

The temperature and speed radial profiles of O$^{6+}$ (see Figure \ref{fig:radial_profiles}) are well organized across wind speed bins.  However no strong speed dependence is found except for O$^{6+}$ temperature. The results indicate that O$^{6+}$ behaves similar to He$^{2+}$ where differential flow and its temperature ratio with protons decreases with heliocentric distance however, unlike alphas, shows no significant heating is experienced between 0.3 to 1au, as indicated by the adiabatic scalar temperature profile.

As evident in Figure \ref{fig:ex1} and \ref{fig:ex2}, O$^{6+}$ exhibits negative differential streaming during periods where large kinks in the magnetic field occur. Detailed examination of the oxygen speed across the magnetic fluctuation indicates its speed is reduced in cases where it is expected to increase, following the behavior of He$^{2+}$ \citep{Goldstein1995, Matteini2014, Matteini2015, McManus2022, Wang2025}. In future work, a detailed and higher resolution examination of the speed of O$^{6+}$ during Alfv\'enic fluctuations at different stages of the solar wind's radial evolution can provide valuable insight to wave-particle dynamics. In particular, careful determination of the local wave speed is key to understanding the behavior of O$^{6+}$ during these fluctuations.

A comparison to remote observations of O$^{5+}$ at the Sun (up to $3.5$ and $5R_{\sun}$, see Figure \ref{fig:compare_UVCS}), suggests the super heated oxygen in the corona may peak in temperature and cool very shortly after the observed distances under the assumption of an adiabatic extrapolation back to the Sun, suggesting the majority of heating happens below 0.3au and the Alfv\'en surface. However, given there is a huge gap between remote observations ending at $5R_{\sun}$ and $65R_{\sun}$ where O$^{6+}$ is first observed in situ, it is unclear where a temperature peak occurs. For a complete picture across this region, there is a critical need for heavy ion observations below 0.3au in both in situ as well as contemporaneous remote observations of heavy ion densities, temperatures, and outflow speeds from a next generation UVCS-like instrument to bridge this gap \citep{Rivera2022, Laming_Lockyer, Rivera2025b}. 

In summation, our observational findings show that the radial evolution of O$^{6+}$, the third most abundant ion species in the solar wind, differs from that of protons and He$^{2+}$ in several key ways. It therefore places new constraints on multi-species models of solar wind heating and acceleration. 

\acknowledgments

The authors thank Benjamin Chandran and Timothy Horbury for insightful discussion on the work. YJR, STB, MLS are partially supported by the Parker Solar Probe project through the SAO/SWEAP subcontract 975569. CJO and DV were supported by STFC Consolidated Grant ST/W001004/1. KGK is partially supported by NASA grant 80NSSC240171. BLA acknowledges Solar Orbiter and Parker Solar Probe funding at NASA/GSFC and NASA/LWS funding under 80NSSC22K1011. JTC was supported by STFC grant ST/W001071/1. JHW is supported by the STFC under studentship ST/X508433/1.

Solar Orbiter is a mission of international cooperation between ESA and NASA, operated by ESA. Solar Orbiter SWA data were derived from scientific sensors that were designed and created and are operated under funding provided by numerous contracts from UKSA, STFC, the Italian Space Agency, CNES, the French National Centre for Scientific Research, the Czech contribution to the ESA PRODEX programme and NASA. SO SWA work at the UCL/Mullard Space Science Laboratory is currently funded by UKSA/STFC grants UKRI919 and UKRI1204. The SWA-HIS team acknowledges NASA contract NNG10EK25C. 

\appendix

\begin{figure}
\centering
\includegraphics[width=0.8\textwidth]{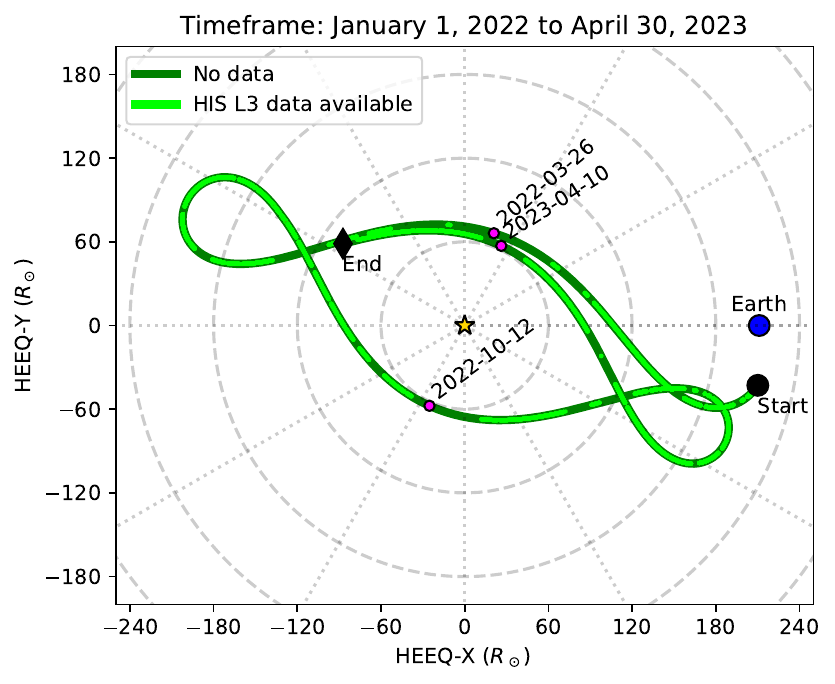}

\caption{The orbital path of Solar Orbiter over the time frame covered in the analysis in this paper is plotted in dark green. The light green periods along the trajectory are locations where the HIS Level 3 data were available. The magenta markers and corresponding dates indicate perihelia. }
\label{fig:context_orbits}
\end{figure}

\section{Solar Orbiter trajectory and data availability} \label{appsec:orbits_traj}
Figure \ref{fig:context_orbits} shows the trajectory of Solar Orbiter within the HIS data availability period of 1 January 2022--30 April 2023. The period includes three perihelia, with closest approaches on 26 March 2022, 12 October 2022, and 10 April 2023.  The light green periods highlighted along the trajectory show periods where HIS data was available. 

\section{Moth plots separated by speed and heliocentric distance} \label{appsec:moth_plots}
\begin{figure*}
\centering

\includegraphics[width=0.31\textwidth]{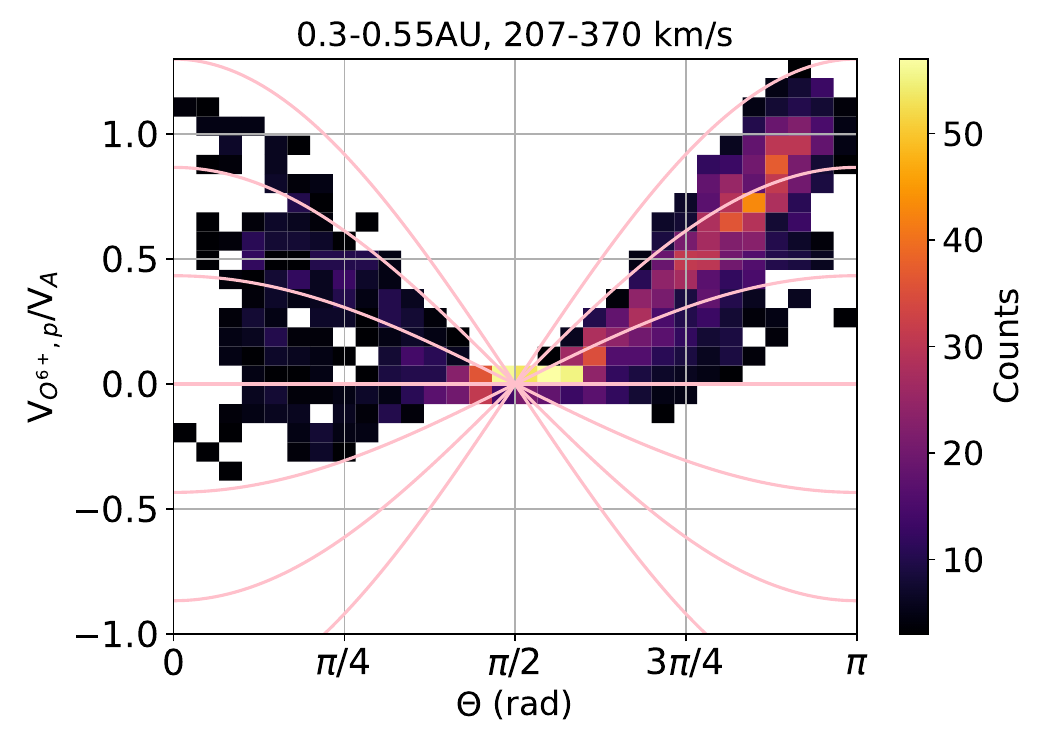}
\includegraphics[width=0.31\textwidth]{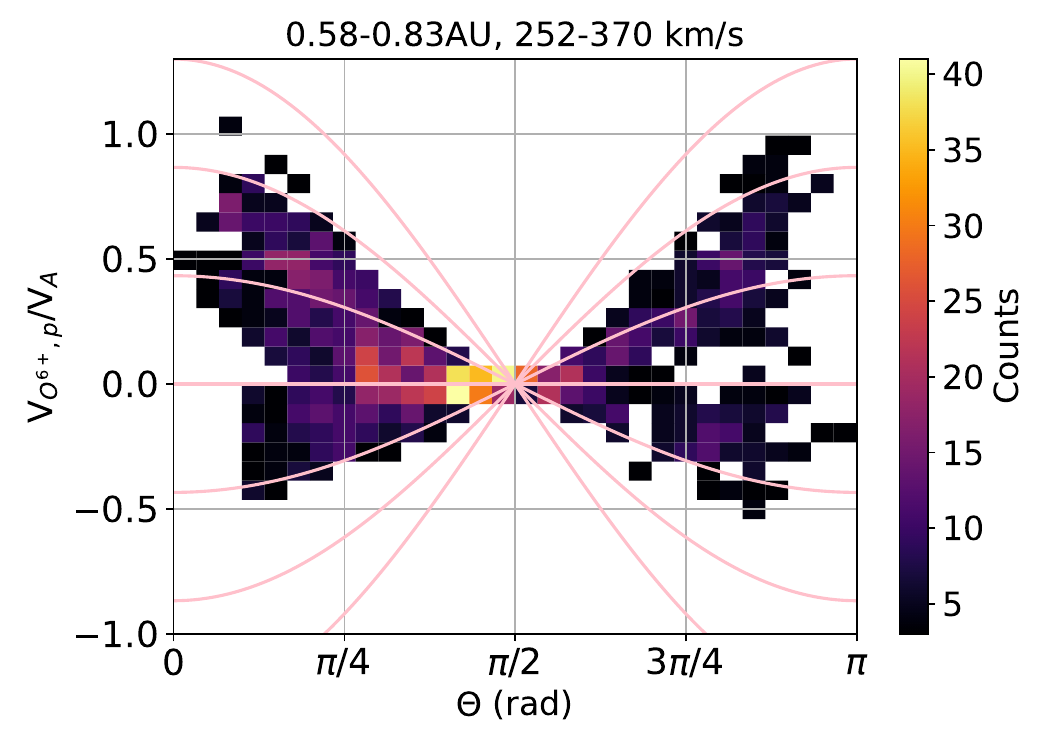}
\includegraphics[width=0.31\textwidth]{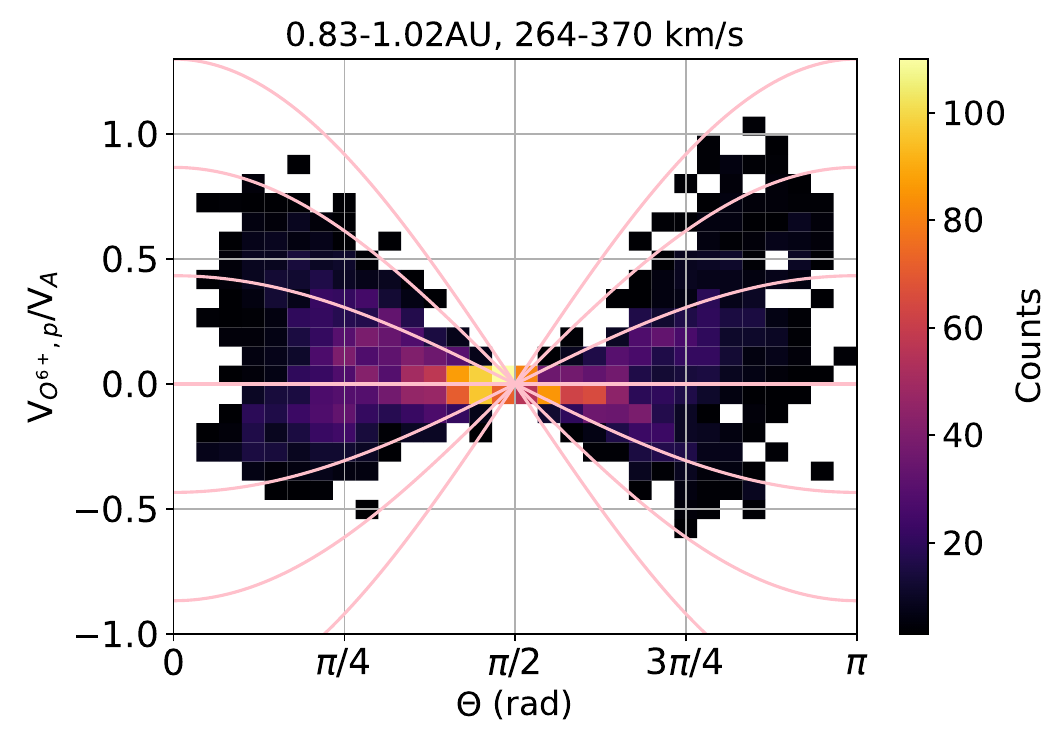}
\includegraphics[width=0.31\textwidth]{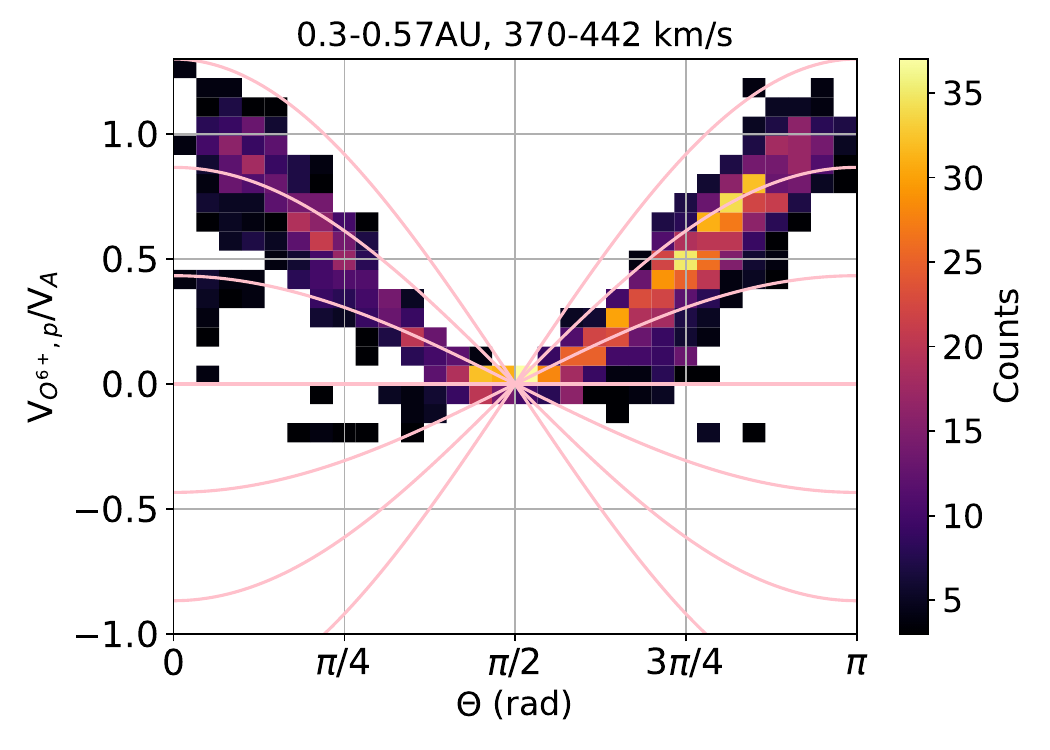}
\includegraphics[width=0.31\textwidth]{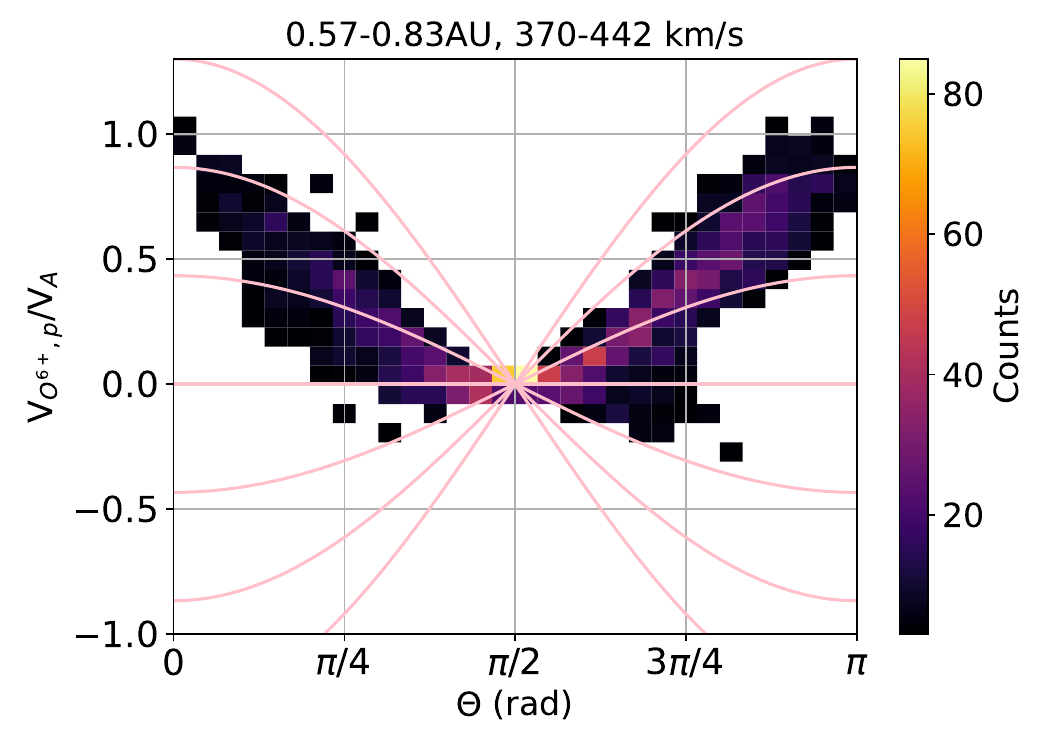}
\includegraphics[width=0.31\textwidth]{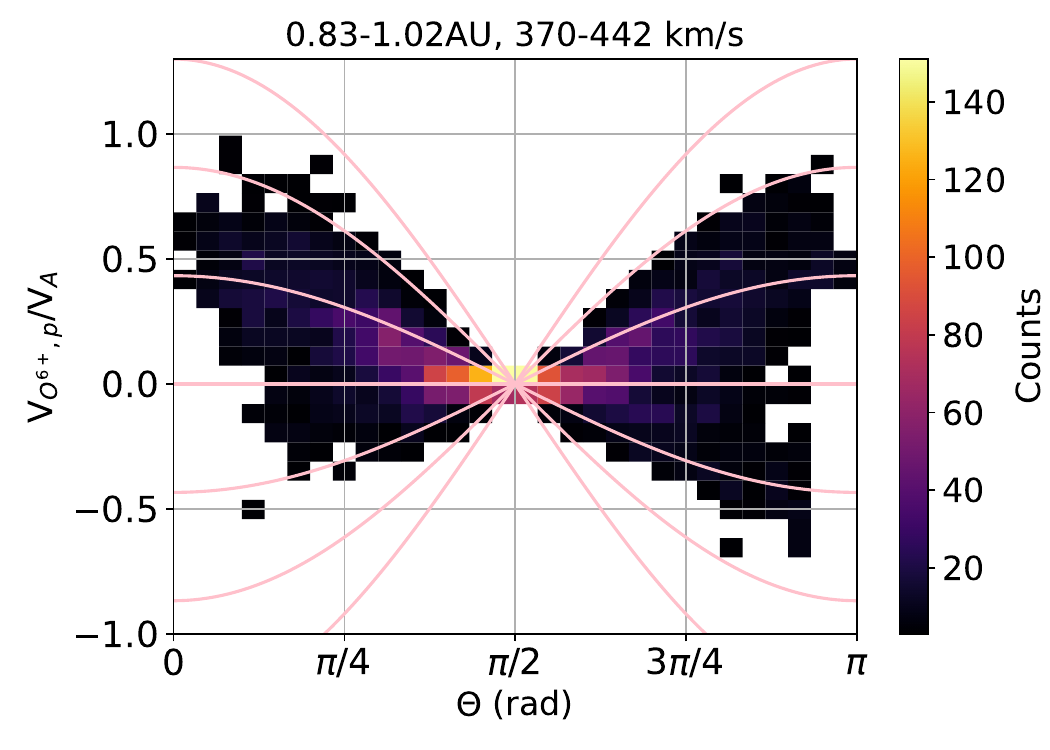}
\includegraphics[width=0.31\textwidth]{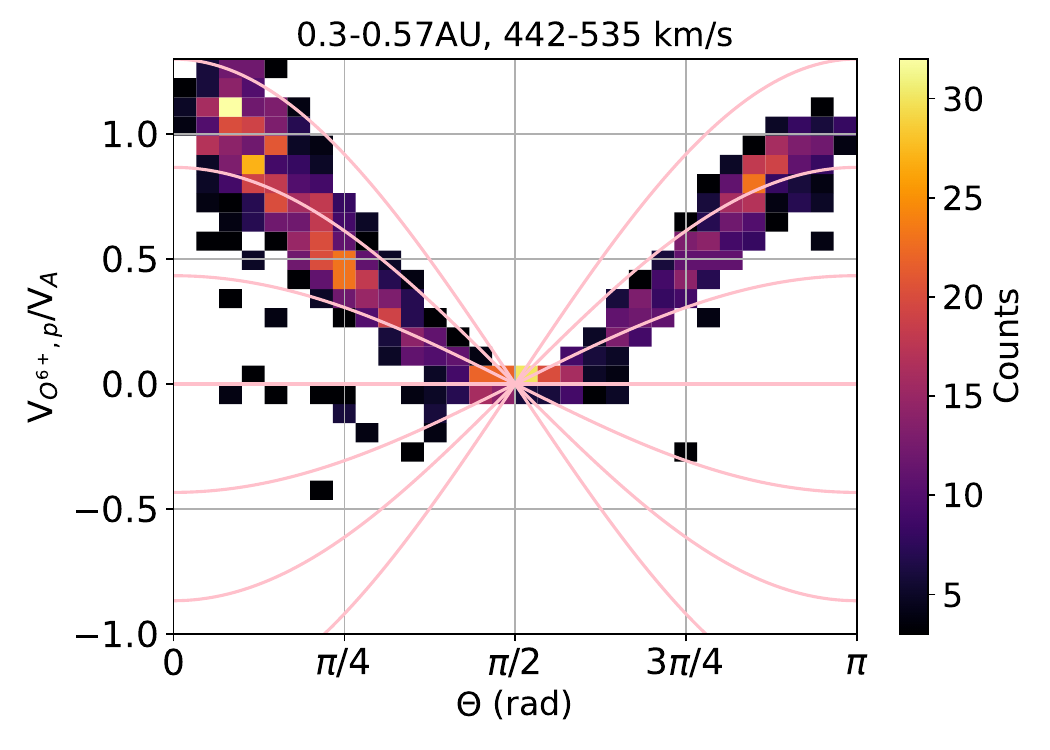}
\includegraphics[width=0.31\textwidth]{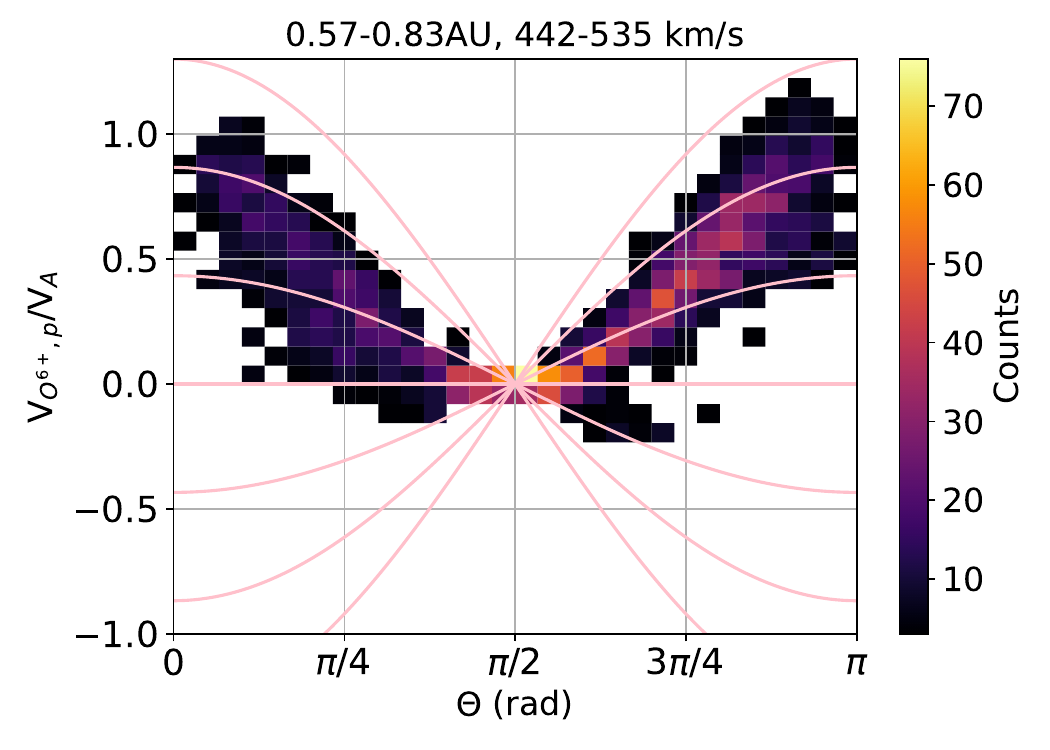}
\includegraphics[width=0.31\textwidth]{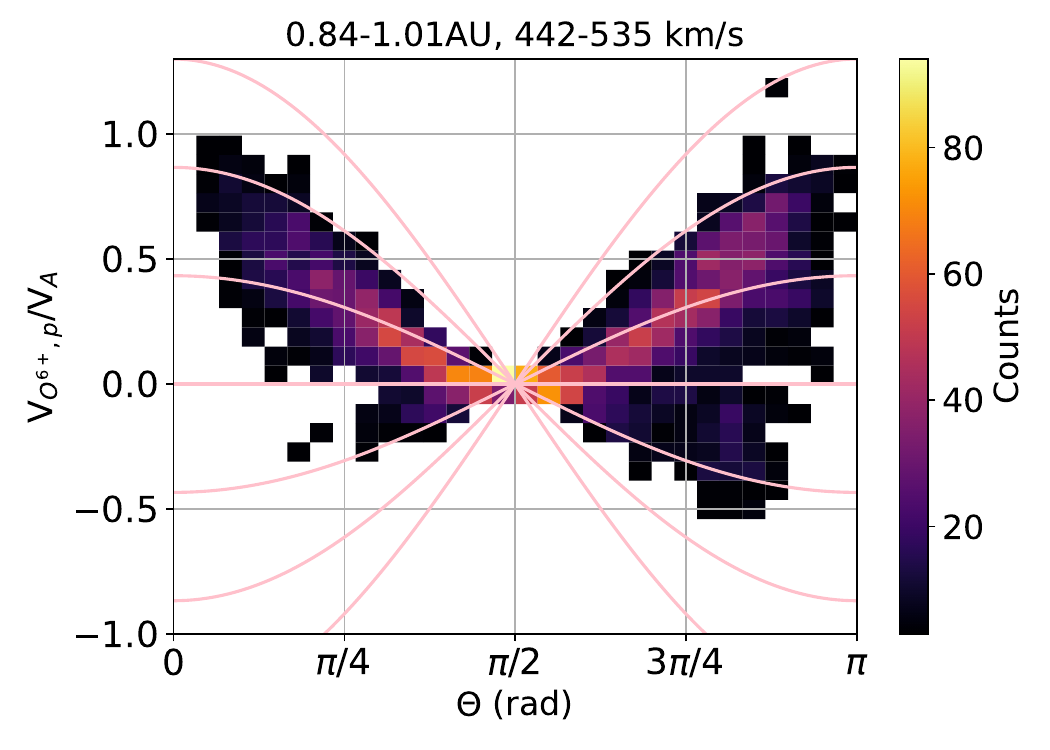}
\includegraphics[width=0.31\textwidth]{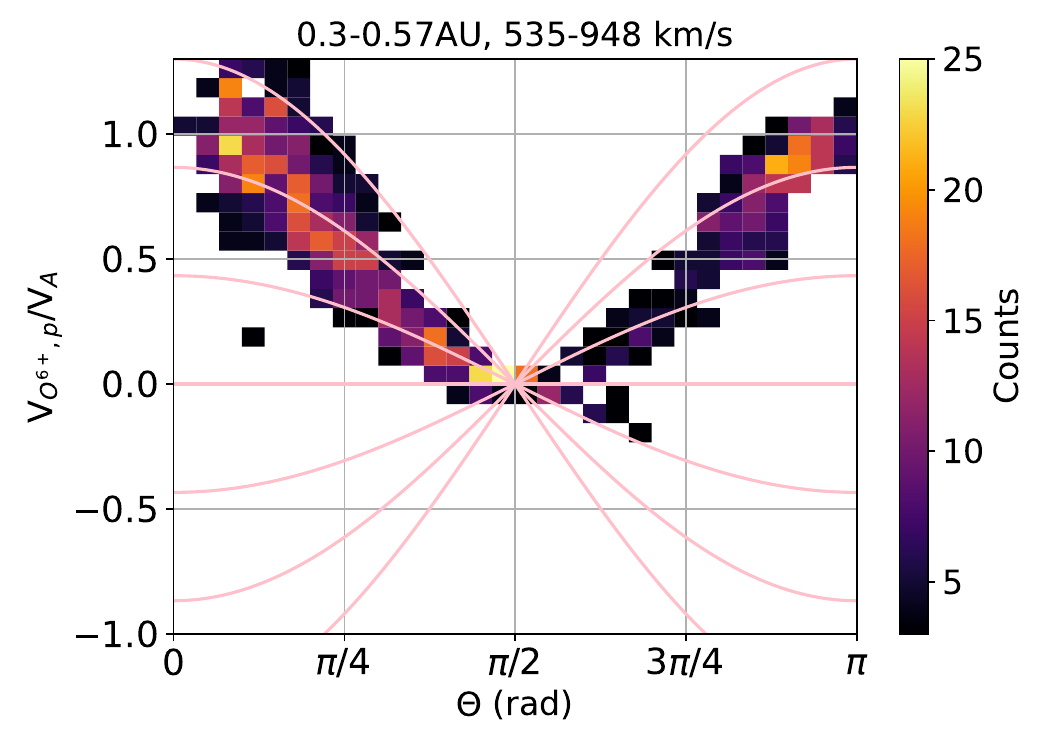}
\includegraphics[width=0.31\textwidth]{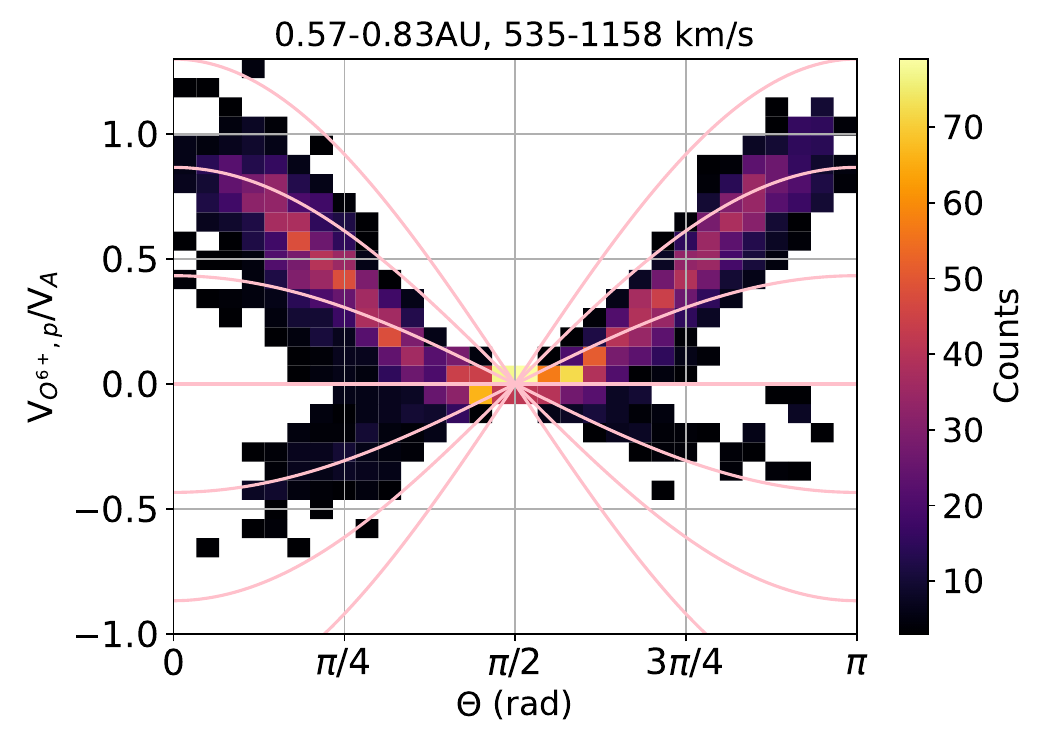}
\includegraphics[width=0.31\textwidth]{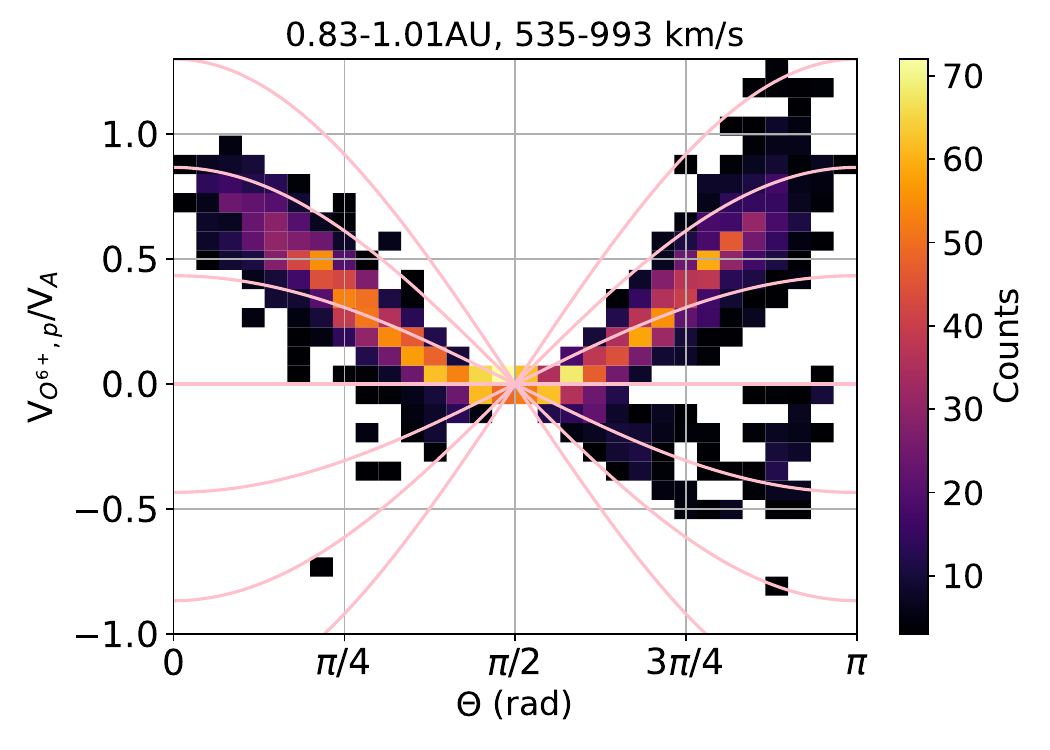}

\caption{Moth plots of counts differential streaming of O$^{6+}$ normalized to the local Alfv\'en speed versus angle between the flow direction and the local magnetic field, same as Figure \ref{fig:differential_streaming_versus_distance_temp}.}
\label{fig:differential_streaming_versus_distance}
\end{figure*}

\begin{figure*}
\centering

\includegraphics[width=0.31\textwidth]{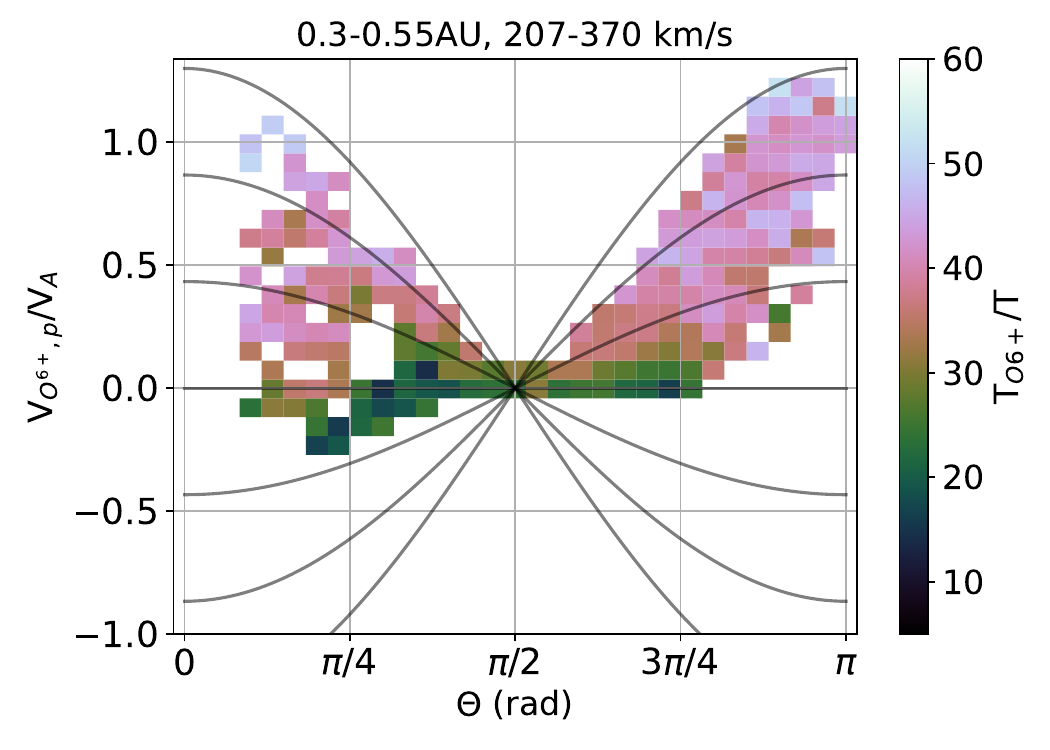}
\includegraphics[width=0.31\textwidth]{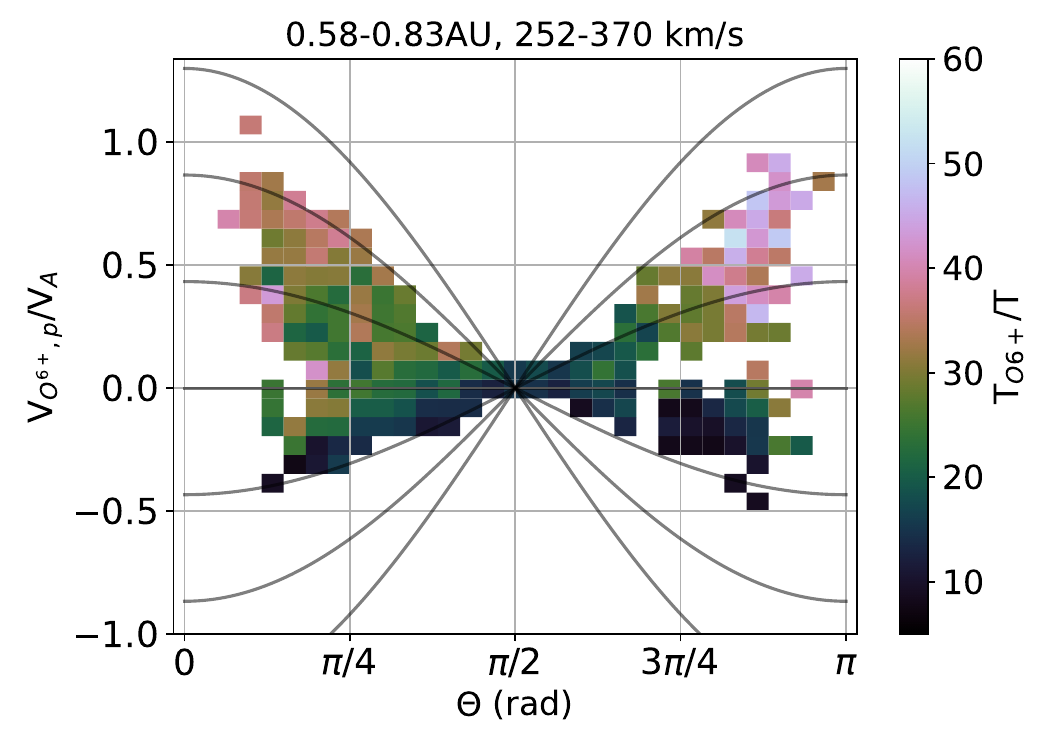}
\includegraphics[width=0.31\textwidth]{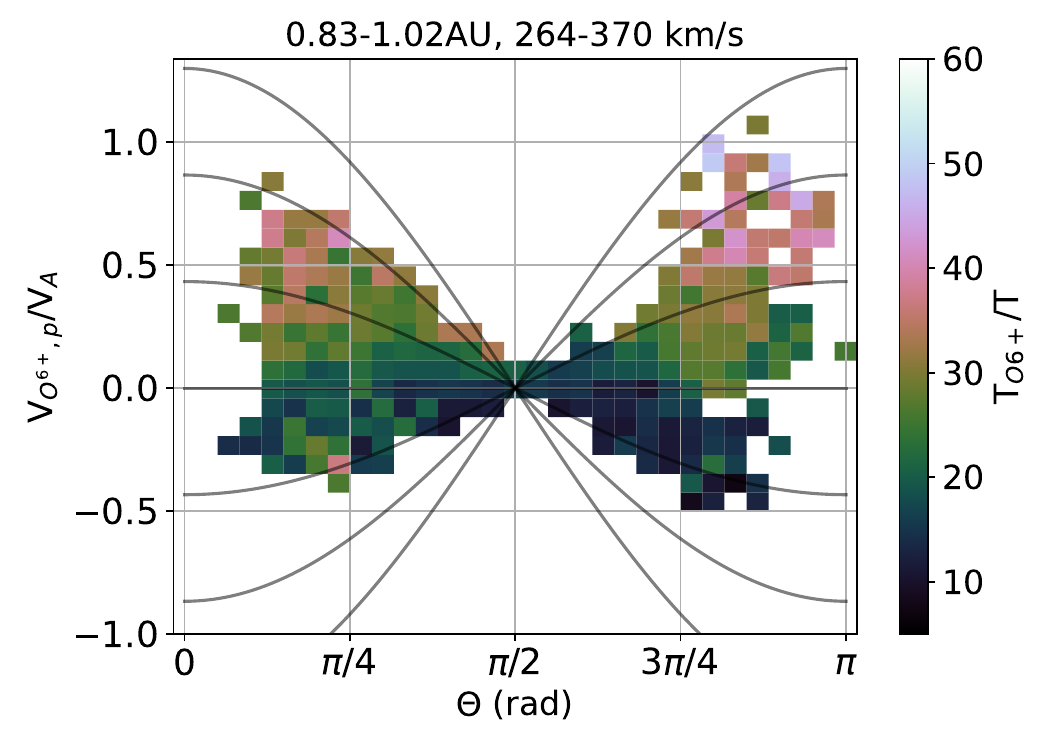}
\includegraphics[width=0.31\textwidth]{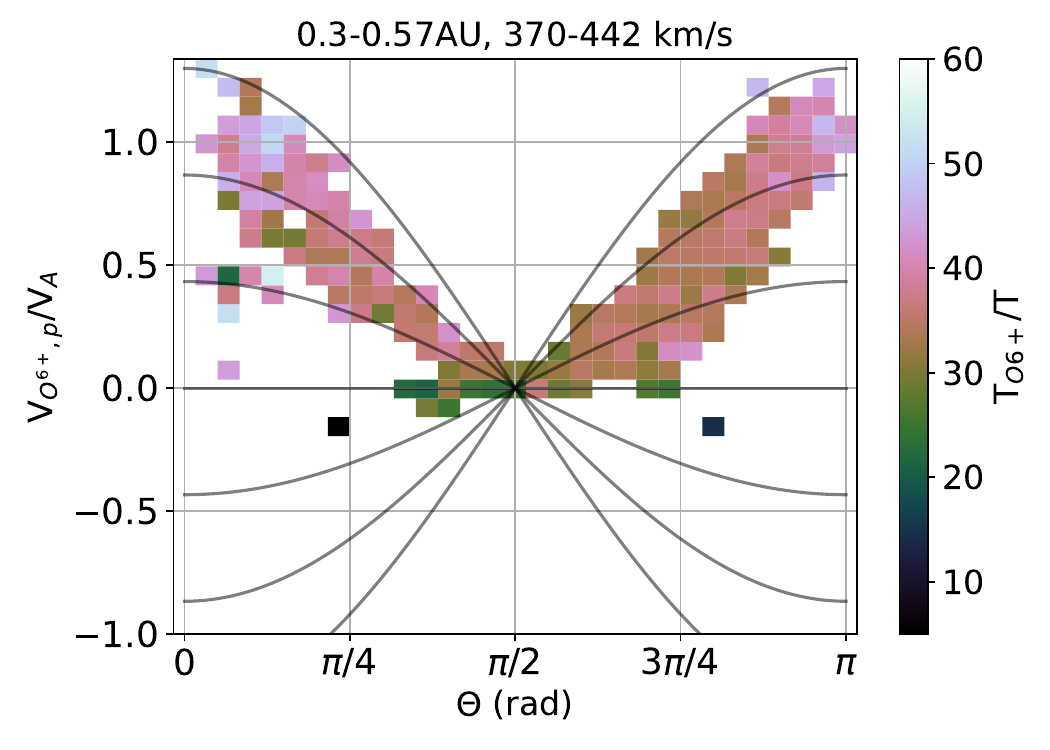}
\includegraphics[width=0.31\textwidth]{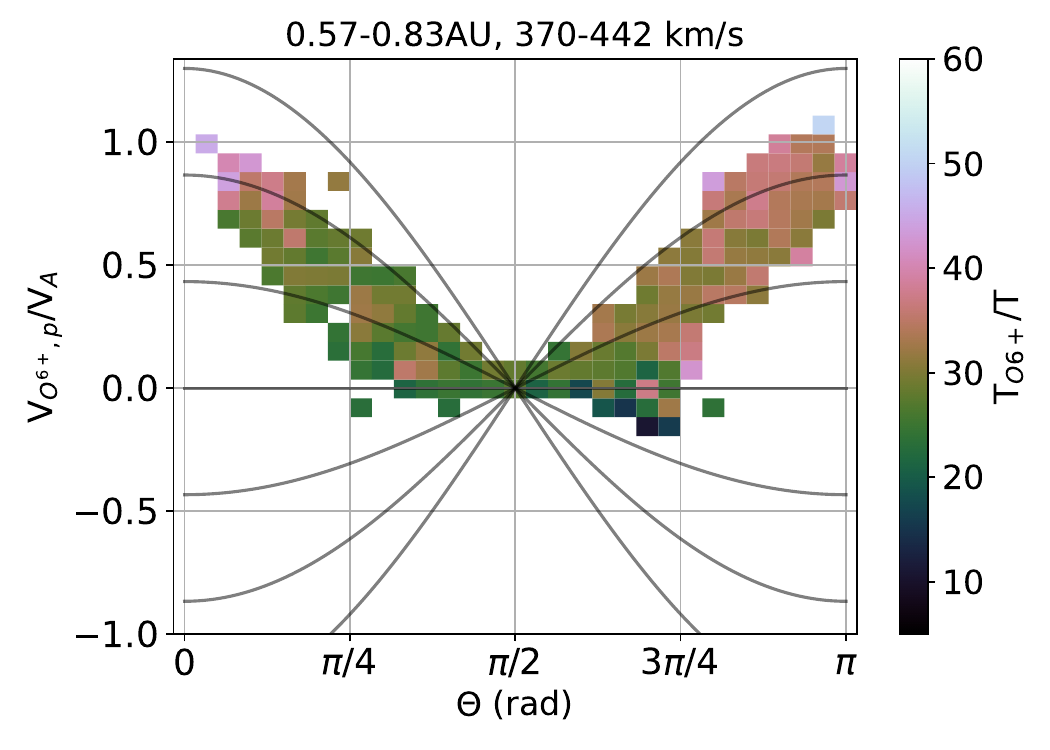}
\includegraphics[width=0.31\textwidth]{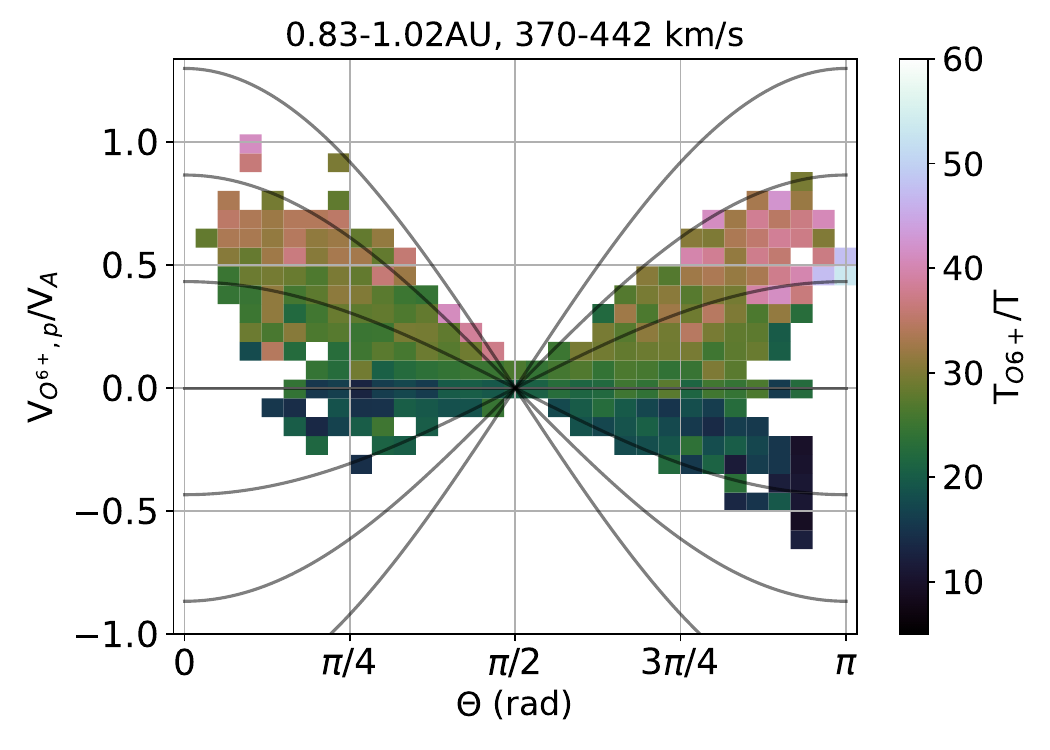}
\includegraphics[width=0.31\textwidth]{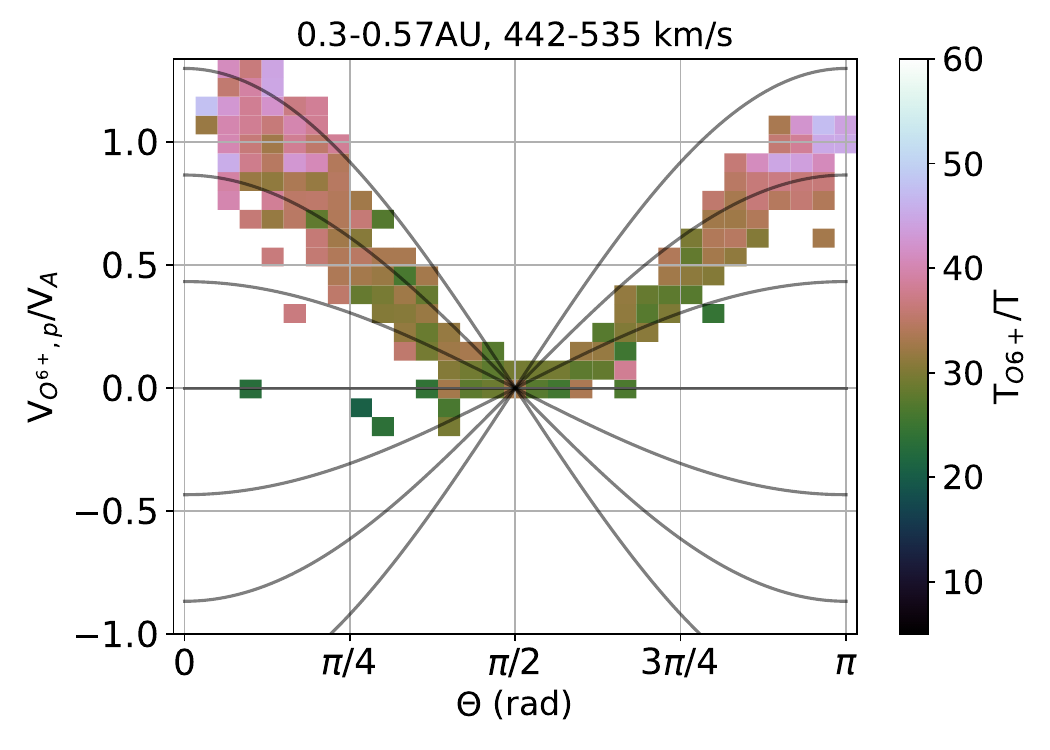}
\includegraphics[width=0.31\textwidth]{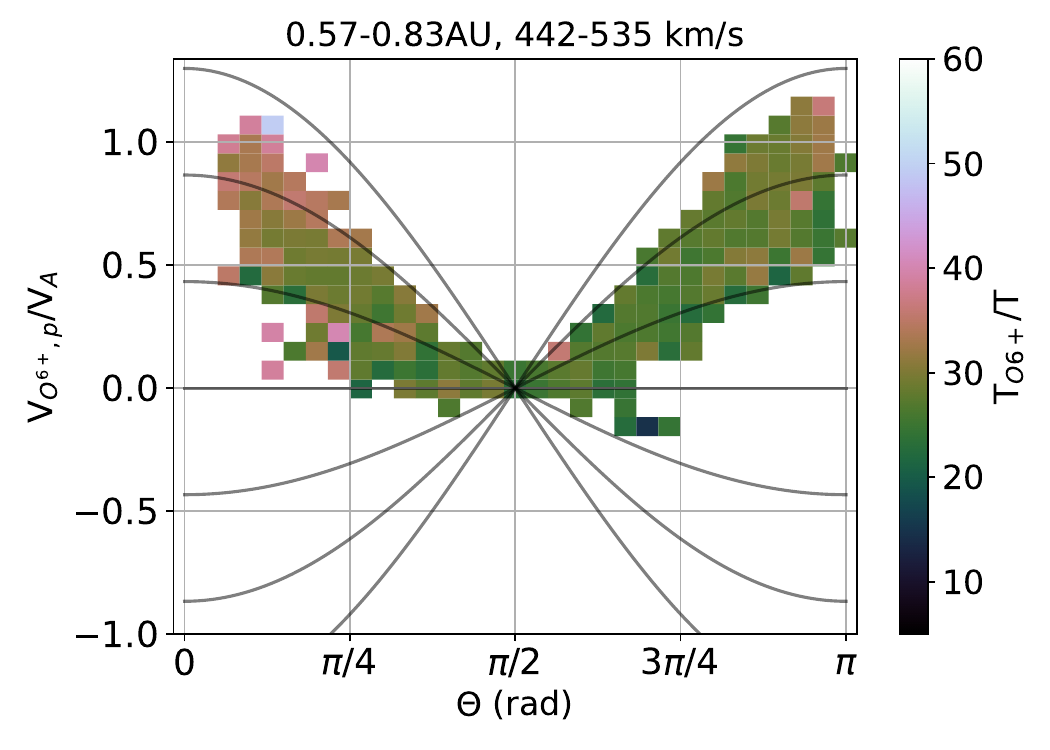}
\includegraphics[width=0.31\textwidth]{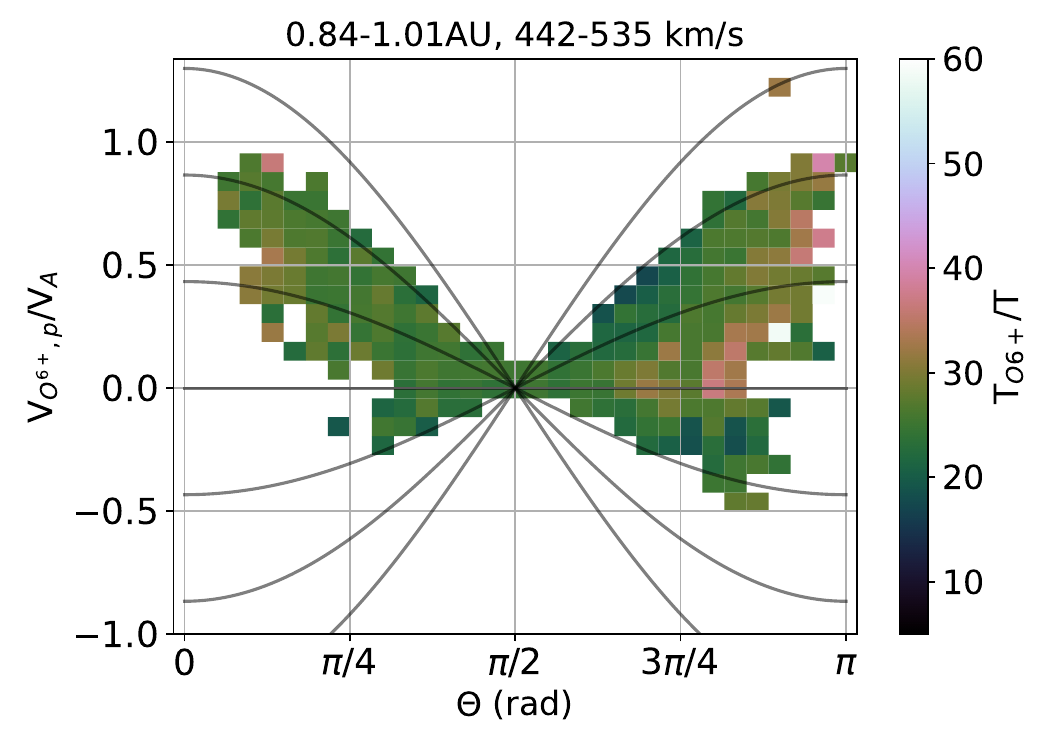}
\includegraphics[width=0.31\textwidth]{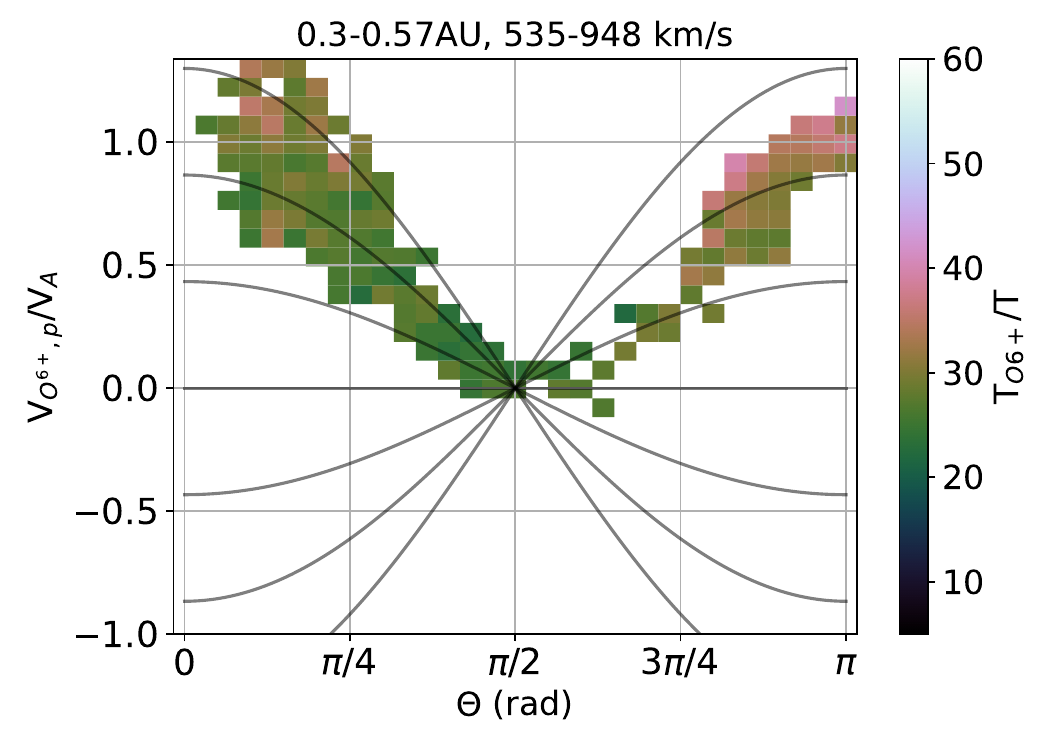}
\includegraphics[width=0.31\textwidth]{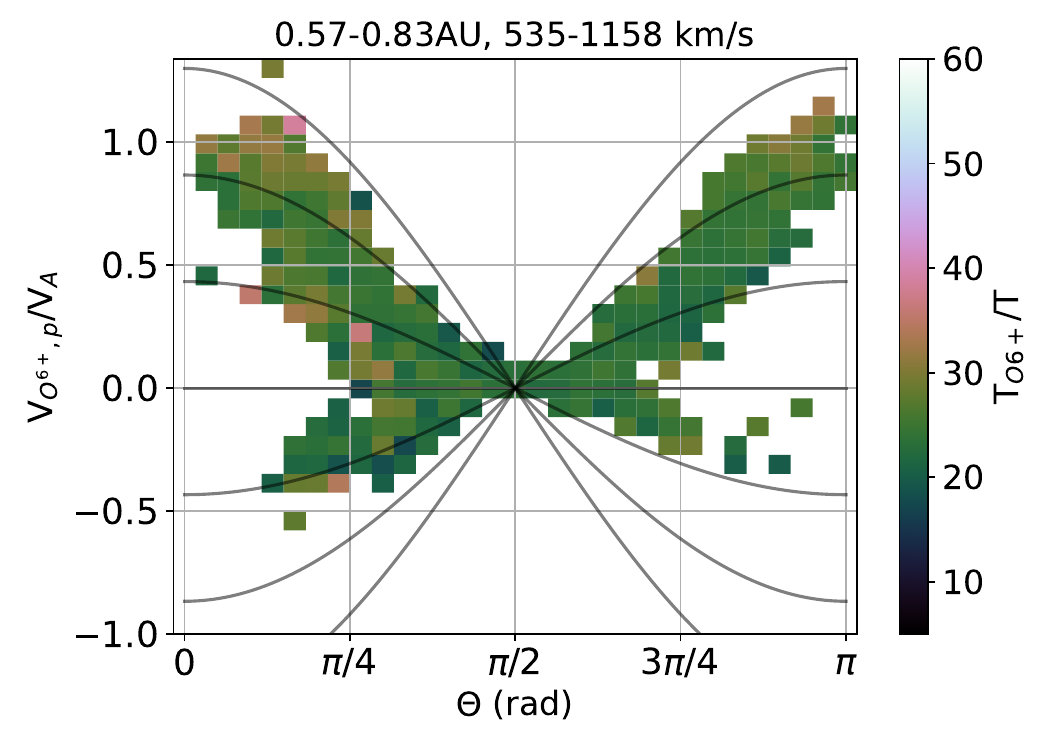}
\includegraphics[width=0.31\textwidth]{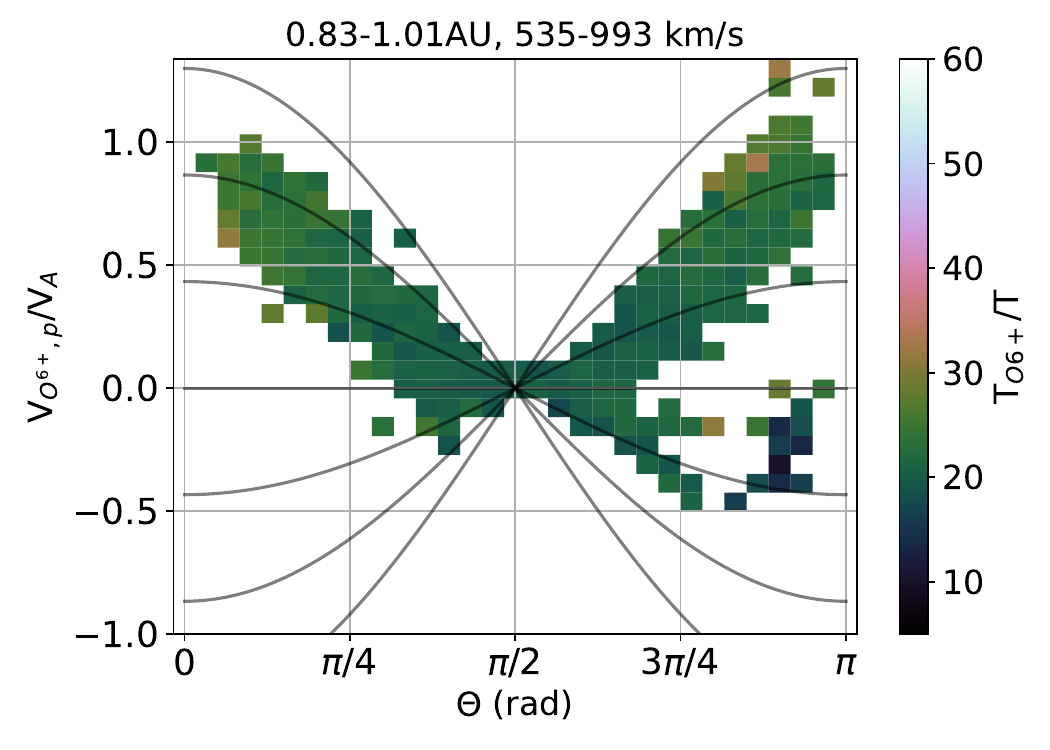}

\caption{Moth plots, 2D histogram, of temperature ratio (T$_{O6+}$/T$_{p}$) for differential streaming of O$^{6+}$ normalized to the local Alfv\'en speed versus angle ($\Theta$) between the radial direction and the local magnetic field direction. The colorbar indicates the proton to O$^{6+}$ temperature ratio (T$_{O6+}$/T$_p$) which is kept consistent across all panels and with Figure \ref{fig:all_differential_streaming}.}
\label{fig:differential_streaming_versus_distance_temp}
\end{figure*}

Figure \ref{fig:differential_streaming_versus_distance} and \ref{fig:differential_streaming_versus_distance_temp} shows the same moth plots as Figure \ref{fig:all_differential_streaming} except computed within various distance intervals from the Sun and in various solar wind speed ranges, as indicated above each panel. Further examining these properties separated by wind speed and distance from the Sun reveals the evolution of the solar wind within this parameter space. The period of interest is divided into 4 speed and 3 distance quantiles to ensure statistically significant comparison between the panels. The ranges of speed and distance noted at the top of each plot are determined from the average data collected in each speed/location range, causing them to differ slightly. The different panels in the figure are arranged with an increasing distance from the Sun viewing left to right and increasing solar wind speed from top to bottom. The colorbar for Figure \ref{fig:differential_streaming_versus_distance} is counts and for Figure \ref{fig:differential_streaming_versus_distance_temp} is the average temperature ratio (T$_{O6+}$/T$_p$) in each bin. We note a few features in the distribution of points, on average: 

\begin{enumerate}
    \item Field-aligned differential streaming is maximum closest to the Sun across all solar wind speeds.
    \item Solar wind with differential streaming larger than an Alfv\'en speed occurs at the closest distances to the Sun.
    \item Differential streaming decreases with distance from the Sun within all solar wind speeds.
    \item Negative differential streaming increases with growing distance from the Sun.
\end{enumerate}

Pertaining to the temperature and differential streaming relationship, we find that the largest temperature ratio is observed closest to the Sun and in the slowest speed bin (top left of Figure \ref{fig:all_differential_streaming} and also in Appendix Figure \ref{fig:differential_streaming_versus_distance_temp}). However, all analyzed solar wind shows a decrease in the temperature ratio with larger heliocentric distances. We discuss this observations further in the following section. The fastest wind speeds (bottom row of Appendix Figure \ref{fig:differential_streaming_versus_distance_temp}) show more uniformity in temperature ratio within changing differential streaming values overall, while the slowest wind contains a larger range of temperature, particularly at large $\Theta$. The negative differential streaming region of the plot is generally lower in temperature ratio, as observed in the right panel of Figure \ref{fig:all_differential_streaming}.

\section{Magnetic and Velocity Fluctuations across Alfv\'enic Fluctuation}
\begin{figure*}
\centering
\includegraphics[width=0.4\textwidth]{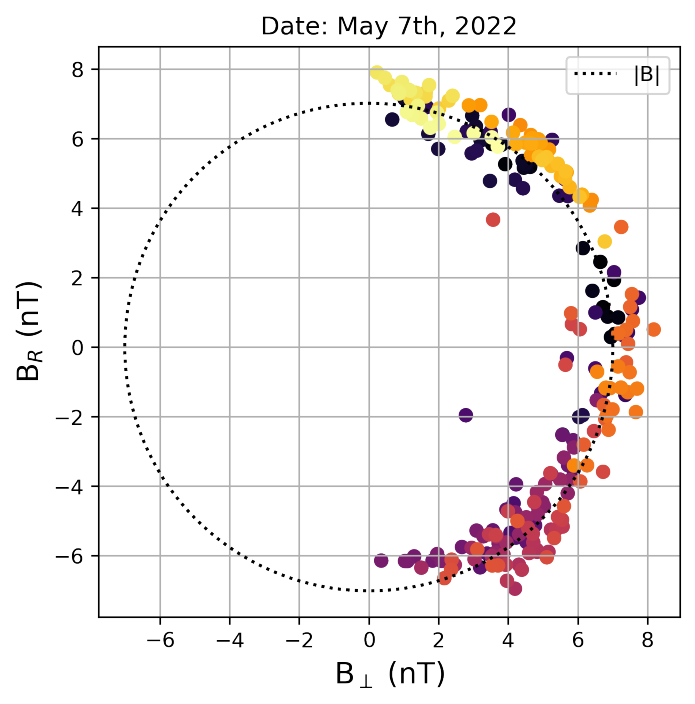}
\includegraphics[width=0.51\textwidth]{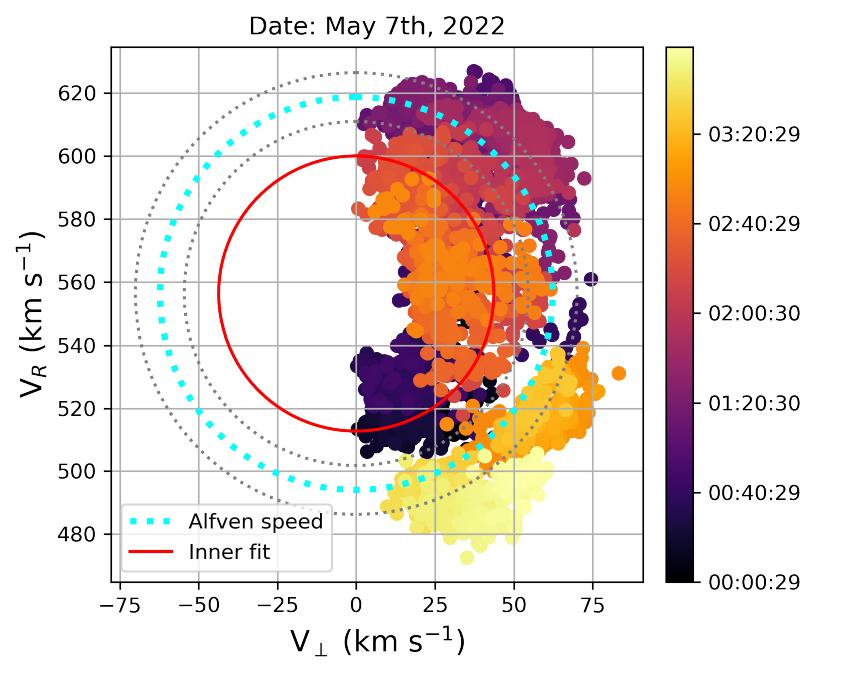}
\caption{The magnetic field and velocity, radial (B$_R$, V$_R$) and perpendicular projections (B$_{\perp}$, V$_{\perp}$) during the shaded magenta region timeframe of Figure \ref{fig:ex1} where the color of the points indicates the associated time.}
\label{fig:radial_perp_B_V}
\end{figure*}

Figure \ref{fig:radial_perp_B_V} shows the magnetic field ($B_R$ and $B_{\perp}$) and velocity ($V_R$ and $V_{\perp}$) components, left and right panels, respectively, across the shaded magenta region of Figure \ref{fig:ex2}, as described in Section \ref{sec:negativediffstream}. The radius of the dashed circle (left panel) is the mean magnitude of the magnetic field, while the cyan circle in the right panel shows the mean V$_A$ and the two black circles are $\pm$ standard deviation of V$_A$. The red circle is a curve fitted to the velocity covering the inner part of the circle, where the V$_A=62$ km s$^{-1}$ (cyan) and the red circle is 43.4 km s$^{-1}$ (red).

\section{Differential streaming versus Alfv\'en speed}
\begin{figure*}
\centering

\includegraphics[width=0.49\textwidth]{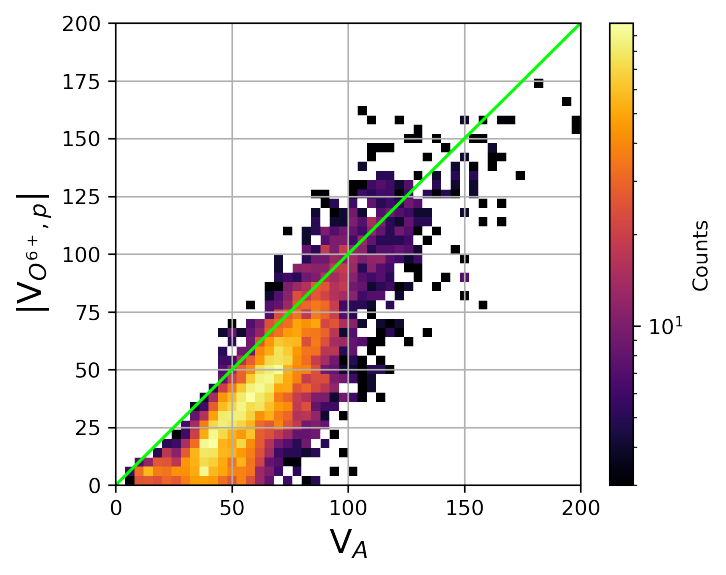}
\includegraphics[width=0.49\textwidth]{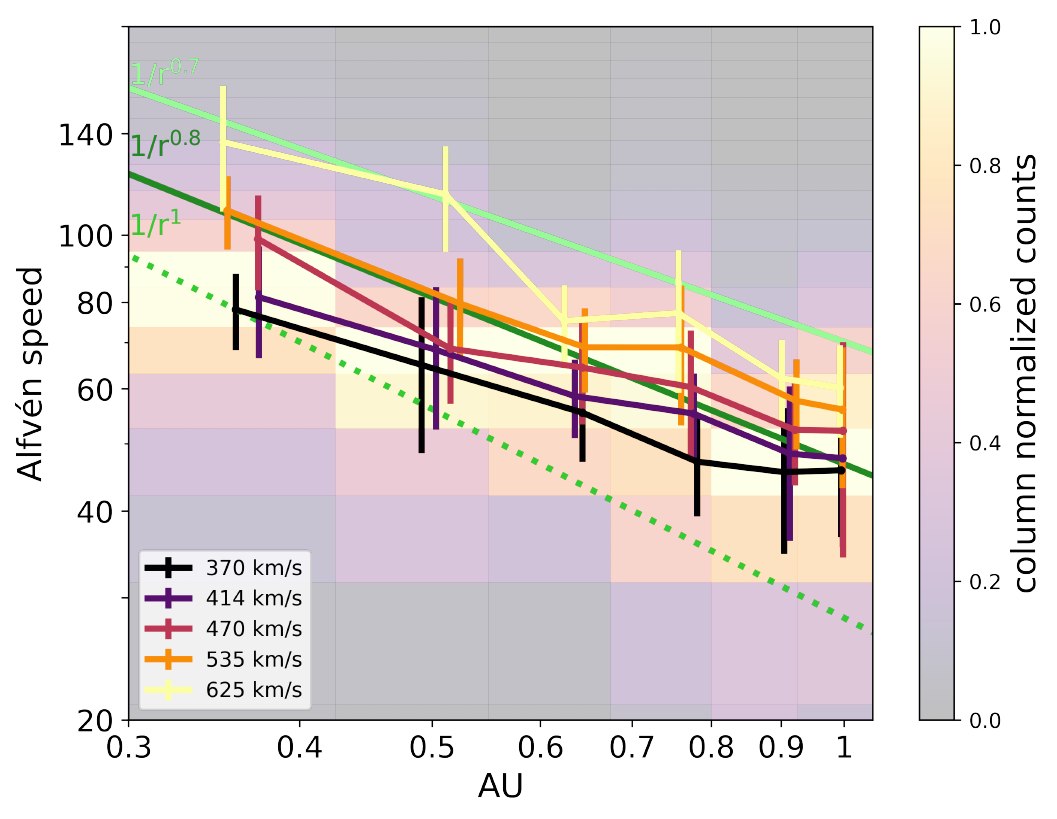}

\caption{(left) 2D histogram of counts of differential speed, V$_{O6+,p}$ against Alfv\'en speed, V$_{A}$. The green line has a slope of 1 showing correlation between the values. (right) Radial evolution of Alfv\'en speed within the same speed bins as \ref{fig:radial_profiles}. We include three power-law profiles: 1/r$^{0.7}$, 1/r$^{0.8}$, 1/r$^1$ for reference.}

\label{fig:diff_vs_Va}
\end{figure*}

Figure \ref{fig:diff_vs_Va}, left panel, shows the distribution of V$_{O6+,p}$ vs Alfv\'en speed. As in the top row of Figure \ref{fig:radial_profiles}, we only include periods of differential streaming for $\Theta <0.5$ and $>2.5$.  The plot indicates a strong positive correlation between differential flow and V$_A$, however does not appear to saturate at the largest Alfv\'en speed as indicated by \cite{Wang2025} for He$^{2+}$. This is likely due to the proximity to the Sun where the large majority of observations from \cite{Wang2025} are taken much closer than 0.3au while Solar Orbiter samples 0.3 and beyond. This is a strong indicator for the need of heavy ion observations below 0.3au where this phenomena occurs for direct comparison to alpha particles.

Figure \ref{fig:diff_vs_Va}, right panel, shows the radial profile of the Alfv\'en speed for the different speed bins from Figure \ref{fig:radial_profiles}. Similar to \cite{Neugebauer1996}, the V$_A$ profiles are shallower than $1/r$ dependence across all wind speeds.

\end{document}